\documentclass[a4paper,11pt]{article}
\usepackage[utf8]{inputenc}

\usepackage{framed}
\usepackage{mdframed}

\usepackage{fullpage}
\usepackage[T1]{fontenc}

\usepackage[british]{babel}
\usepackage{verbatim}
\usepackage[T1]{fontenc}
\usepackage{lmodern}
\usepackage{lipsum}
\usepackage{booktabs}
\usepackage{caption}
\usepackage{cite}
\usepackage{soul,color}
\usepackage[toc,page]{appendix}

\usepackage{ifpdf}

\usepackage{amsthm, amssymb}
\usepackage[tbtags]{amsmath}
\usepackage{bm}
\usepackage{mathtools}
\usepackage{amstext}
\usepackage{braket}
\usepackage{multirow}

\usepackage[normalem]{ulem}
\usepackage{xcolor}

\newcommand\redsout{\bgroup\markoverwith{\textcolor{red}{\rule[0.5ex]{2pt}{0.4pt}}}\ULon}

\begin{document}
\vspace{5mm}
\vspace{0.5cm}

\def\be{\begin{eqnarray}}
\def\ee{\end{eqnarray}}

\def\ba{\begin{aligned}}
\def\ea{\end{aligned}}

\def\ls{\left[}
\def\rs{\right]}
\def\lc{\left\{}
\def\rc{\right\}}

\def\p{\partial}

\def\S{\Sigma}

\def\s{\sigma}

\def\O{\Omega}

\def\a{\alpha}
\def\b{\beta}
\def\g{\gamma}

\def\ad{{\dot \alpha}}
\def\bd{{\dot \beta}}
\def\gd{{\dot \gamma}}
\newcommand{\ft}[2]{{\textstyle\frac{#1}{#2}}}
\def\ib{{\overline \imath}}
\def\jb{{\overline \jmath}}
\def\Re{\mathop{\rm Re}\nolimits}
\def\Im{\mathop{\rm Im}\nolimits}
\def\trace{\mathop{\rm Tr}\nolimits}
\def\rmi{{ i}}

\def\N{\mathcal{N}}

\newcommand{\SU}{\mathop{\rm SU}}
\newcommand{\SO}{\mathop{\rm SO}}
\newcommand{\U}{\mathop{\rm {}U}}
\newcommand{\USp}{\mathop{\rm {}USp}}
\newcommand{\OSp}{\mathop{\rm {}OSp}}
\newcommand{\Symp}{\mathop{\rm {}Sp}}
\newcommand{\Sl}{\mathop{\rm {}S}\ell }
\newcommand{\Gl}{\mathop{\rm {}G}\ell }
\newcommand{\Spin}{\mathop{\rm {}Spin}}

\def\hc{c.c.}

\numberwithin{equation}{section}

\allowdisplaybreaks

\allowbreak



\begin{titlepage}
	\thispagestyle{empty}
	\begin{flushright}

	\end{flushright}
\vspace{35pt}

	\begin{center}
	    { \Large{
	    Supersymmetric Born--Infeld actions and new Fayet--Iliopoulos terms 
	     }}

		\vspace{50pt}

		{Niccol\`o~Cribiori$^{1}$, Fotis~Farakos$^{2}$ and Magnus~Tournoy$^{2}$ }

		\vspace{25pt}

		{
			$^1$ {\it Institute for Theoretical Physics, TU Wien,\\ Wiedner Hauptstrasse 8-10/136, A-1040 Vienna, Austria }

		\vspace{15pt}

			$^2${\it   KU Leuven, Institute for Theoretical Physics, \\
			Celestijnenlaan 200D, B-3001 Leuven, Belgium}
            
 		}

		\vspace{40pt}

		{ABSTRACT}
	\end{center}

	\vspace{10pt}

We consider ${\cal N}=1$ supersymmetric Born--Infeld actions that have a second non-linear supersymmetry. We focus on the model proposed by Bagger and Galperin and show that the breaking of the second supersymmetry is sourced by the new Fayet--Iliopoulos D-term. Interpreting such an action as the effective theory of a space-filling (anti) D3-brane leads to an expression for the new Fayet--Iliopoulos parameter in terms of the brane tension and $\alpha'$.

\bigskip

\end{titlepage}


\def\thefootnote{\fnsymbol{footnote}}

\vskip 0.5cm

\vspace{0.5cm}

\def\thefootnote{\arabic{footnote}}
\setcounter{footnote}{0}

\baselineskip 5.6 mm






\section{Introduction}

Supersymmetry breaking can be generated with a diversity of methods in four-dimensional supergravity \cite{FVP}, but to understand which of the resulting models might descend from string theory is non-trivial: not all the supergravity constructions are expected to share a connection with the physics in the high energy regime. One of the most studied mechanisms for spontaneous supersymmetry breaking in four dimensions is D-term breaking. The prototype example in global supersymmetry is due to Fayet and Iliopoulos \cite{Fayet:1974jb}, while the non-trivial extension to supergravity has been constructed by Freedman \cite{Freedman:1976uk}. The presence of a Fayet--Iliopoulos D-term in $\mathcal{N}=1$ supergravity requires the existence of a local $U(1)$ R-symmetry, which restricts the allowed interactions \cite{Barbieri:1982ac,Villadoro:2005yq} and hinders a connection with string theory \cite{Komargodski:2009pc,Dienes:2009td}.

A new embedding of the Fayet--Iliopoulos D-term in supergravity has been constructed in \cite{Cribiori:2017laj}. It does not require the gauging of the R-symmetry and thus avoids the aforementioned restrictions. 
For instance, the {\it no-go} theorem of \cite{Komargodski:2009pc} does not apply, leaving room for a string theory interpretation. 
The couplings arising from \cite{Cribiori:2017laj} share similarities with those coming from non-linear realizations of local supersymmetry 
\cite{Lindstrom:1979kq,Samuel:1982uh,Antoniadis:2014oya,Ferrara:2014kva,Kallosh:2014via,DallAgata:2014qsj,Ferrara:2016een,Bandos:2016xyu} 
and have been employed in cosmological models in supergravity 
\cite{Aldabergenov:2017hvp,Antoniadis:2018cpq,Farakos:2018sgq,Antoniadis:2018oeh,Antoniadis:2018ngr,Aldabergenov:2018nzd}. 
In particular, once chiral matter superfields are introduced, the impact of the new Fayet--Iliopoulos D-term on the scalar potential matches the uplift term induced by an anti D3-brane at strong warping \cite{Kachru:2003aw,Kachru:2003sx}, which is commonly described, within four-dimensional supergravity, by non-linear realizations and constrained superfields \cite{Ferrara:2014kva,Kallosh:2014via,Bandos:2016xyu}. 
These considerations seem to suggest that there might be a string theory origin of the new Fayet--Iliopoulos D-term. In this work we wish to strengthen the interpretation of the new Fayet--Iliopoulos D-term of \cite{Cribiori:2017laj} as an effective description of a space-filling anti D3-brane, within a four-dimensional $\mathcal{N}=1$ supergravity setup.

Our starting point is global $\mathcal{N}=1$ supersymmetry, where we first revisit the supersymmetrization of the Born--Infeld  action. Generic $\mathcal{N}=1$ supersymmetric Born--Infeld actions have been constructed in \cite{Cecotti:1986gb} and they take the form 
\begin{equation} 
\label{BIFC}
S_\text{super BI} \sim \int d^4 x  \, \left( d^2 \theta \, W^2  + c.c. \right)+ \int d^4 x \, d^4 \theta  \, W^2 \overline W^2 \, \Psi \left( W, DW, \ldots \right) \, , 
\end{equation}
with $W_\alpha = -\frac14 \overline D^2 D_\alpha V$, where $V$ is an $\mathcal{N}=1$ vector superfield. The function $\Psi$ in \eqref{BIFC} depends on $W_\alpha$, $\overline W_{\dot \alpha}$ and on their superspace derivatives. Bagger and Galperin have shown in \cite{Bagger:1996wp} that, for a specific form of the function $\Psi$, which we will call $\Psi_\text{BG}$, the action \eqref{BIFC} has a second supersymmetry non-linearly realized and also enjoys an electric-magnetic duality invariance \cite{Bagger:1996wp}. 
These properties suggest a relation with the effective action of a space-filling D3-brane \cite{Rocek:1997hi,Kallosh:2014wsa,Vercnocke:2016fbt,Kallosh:2016aep, Bergshoeff:2015jxa}. 
The spectrum of a D3-brane \cite{Cederwall:1996pv,Aganagic:1996pe,Aganagic:1996nn,Bergshoeff:1996tu} contains indeed a vector multiplet, but also a triplet of chiral multiplets. The latter can however be truncated, leaving only the former in the low energy effective description. In this perspective, the Bagger--Galperin action can be interpreted as an effective action for a D3-brane.

The Bagger--Galperin action \cite{Bagger:1996wp} is constructed on a flat four-dimensional background. It can describe therefore both a (truncated) D3 or an anti D3-brane. Indeed, the D3 and anti D3 actions match on a Minkowski background in the $\kappa$-symmetry gauge where the Wess--Zumino term vanishes \cite{Aganagic:1996nn}. When four-dimensional $\mathcal{N}=1$ supergravity is switched-on, however, the two actions should differ, as the supergravity background will respect only one of the two supersymmetries of the (anti) D3-brane action. In particular, when the Bagger--Galperin action is coupled to $\mathcal{N}=1$ supergravity, the linear supersymmetry is preserved, while the second non-linear supersymmetry is explicitly broken.

In this work we show that the Bagger--Galperin action can be presented in an alternative superspace form, which is still of the type \eqref{BIFC}, but with a function $\Psi$ different from the $\Psi_\text{BG}$ of \cite{Bagger:1996wp}. 
We will refer to it as $\Psi_{\overline{\text{BG}}}$. 
In this alternative formulation, the action has the spontaneously broken supersymmetry manifest and described by superspace, while the unbroken supersymmetry is hidden and acquires a complicated form. 
In contrast to the formulation with $\Psi_\text{BG}$, once we insert $\Psi_{\overline{\text{BG}}}$ in \eqref{BIFC} and couple to $\mathcal{N}=1$ supergravity, we find that the spontaneously broken supersymmetry is gauged, whereas the other one is explicitly broken. 
This supports the interpretation of the alternative form of the Bagger--Galperin action that we present in this work as the effective action of an anti D3-brane.

Once we investigate the source of the supersymmetry breaking in the alternative form of the action, we find that it corresponds to the new Fayet--Iliopoulos D-term of \cite{Cribiori:2017laj}. In such an alternative description, therefore, the auxiliary field of the vector multiplet gets a non-vanishing vacuum expectation value and breaks supersymmetry spontaneously. 
Interpreting the Bagger--Galperin action as the action for the truncated (anti) D3-brane on a flat background, its bosonic sector is 
\begin{equation}
{S}^{\text{bos}}\sim- \text{T} \int d^4x\, \sqrt{-\det\left(\eta_{mn} + 2 \pi \alpha' F_{mn}\right)} \, , 
\end{equation}
where T is the brane tension. As we will show in this article, the new Fayet--Iliopoulos parameter is then given by 
\begin{equation}
\xi_\text{new FI} =  2  \pi \alpha' \text{T} \, . 
\end{equation}

Our work is organized as follows. In the next section we review the Bagger--Galperin model and we present the alternative form of its action, in which the non-linear supersymmetry is manifest. In the third section we elaborate on the weak-field limit, which gives rise to the new Fayet--Iliopoulos D-term, and we discuss the differences with the standard Fayet--Iliopoulos term. In the fourth section we couple the alternative action to supergravity and in the fifth section we draw our conclusions. Throughout the article there are a number of technical passages which we have reserved for the appendix. These technical parts are essential for our results, however we have presented them in Appendix \ref{appA} to avoid cluttering the main text with the intricate formalism of non-linear realizations. 
In Appendix B we discuss the deformation of the Bagger--Galperin action with the standard Fayet--Iliopoulos term. 
Throughout this work we use superspace and the conventions of \cite{Wess:1992cp}, but we reserve the Appendix C for a short presentation in the tensor calculus setup of \cite{FVP}.

\section{The Bagger--Galperin action}

The embedding of Born--Infeld actions in $\mathcal{N}=1$ supersymmetry was presented in \cite{Cecotti:1986gb}. A subclass of these actions has a second supersymmetry non-linearly realized, as was derived in \cite{Bagger:1996wp} by Bagger and Galperin, who discussed also the possible relation with D3-branes. In this section we revisit the Bagger--Galperin action. We start from the known formulation in terms of a linear representation of supersymmetry, namely an $\mathcal{N}=1$ vector superfield, and then we recast the model into an equivalent form, in which the spontaneously broken supersymmetry is manifest and described by superspace. In this alternative formulation the supersymmetry breaking is sourced by the new Fayet--Iliopoulos D-term of \cite{Cribiori:2017laj}. As an intermediate step, we obtain a Born--Infeld Lagrangian in which the goldstino sector is described by a Volkov--Akulov fermion \cite{Volkov:1973ix,Ivanov:1978mx,Ivanov:1982bpa,Klein:2002vu,Clark:2005qu} 
and supersymmetry is manifestly non-linearly realized \cite{Bellucci:2015qpa}.

\subsection{The Bagger--Galperin action with manifest non-linear supersymmetry}

The basic ingredient in order to formulate the Bagger--Galperin action in superspace is an $\mathcal{N}=1$ abelian vector superfield $V$.\footnote{The generalization to an arbitrary number of vector fields has been studied recently in \cite{Ferrara:2014oka}, where the relation with the underlying $\mathcal{N}=2$ special geometry is investigated. In \cite{Cribiori:2018jjh} the Bagger--Galperin model is formulated by means of three-forms multiplets. A superembedding formulation is presented in \cite{Bandos:2001ku}.} It has the $\theta$-expansion 
\begin{equation}
\label{Vexp}
V = \ldots - \theta \sigma^m \overline \theta v_m 
+ i \theta^2 \overline \theta \overline \chi 
- i  \overline \theta^2  \theta \chi 
+ \frac12 \theta^2 \overline \theta^2 \text{D} \,  ,
\end{equation}
with the lower dots standing for leading order terms in $\theta$, which are not independent component fields, but depend on the gauge choice. In the standard Wess--Zumino gauge, for example, they are set to vanish \cite{Wess:1992cp}. 
The physical component fields are therefore the abelian gauge vector $v_m$ and the gaugino $\chi_\alpha$, while D is a real scalar auxiliary field. The field strength of the abelian gauge vector $v_m$ resides in the chiral superfield 
\begin{equation}
\label{WV}
W_\alpha = -\frac14 \overline D^2 D_\alpha V \, , 
\end{equation}
which has the standard mass dimension of a fermion in four dimensions, namely $3/2$. The Bagger--Galperin Lagrangian has the form 
\begin{equation}
\label{BG1}
{\cal L}_\text{BG} =  \beta \int d^2 \theta \, {\cal X} + \hc \, , 
\end{equation}
where $\beta$ is a constant with mass dimension 2 and ${\cal X}$ is a nilpotent $\N=1$ chiral superfield, which is a function of $W_\alpha$ defined by 
\begin{equation}
\label{NILP}
{\cal X}= \frac{W^2}{4m + \overline D^2 \overline{\cal X} } \, . 
\end{equation}
The constraint \eqref{NILP} is solved recursively and the explicit expression of $\mathcal{X}$ in terms of $W_\alpha$ is 
\begin{equation}
{\cal X} = \frac{1}{4 m}\left[W^2-\frac12 \overline D^2\left(\frac{W^2\overline W^2}{4m^2+ \frac12 A_+ 
+\sqrt{16m^4+4m^2A_++\frac12 A_-^2}}\right)\right] \, . 
\end{equation}
The constant $m$ has mass dimension 2, while $A_+$ and $A_-$ are defined as
\begin{equation}
A_+ = \frac12 \left(D^2 W^2+\overline D^2\overline W^2\right) \, , \quad  
A_- = \frac12 \left(D^2 W^2-\overline D^2\overline W^2\right) \, . 
\end{equation}
Notice that the Lagrangian \eqref{BG1} is indeed of the type \eqref{BIFC} and therefore it has a manifest linear supersymmetry, namely the one described by superspace, which is preserved. In addition there is a second, non-manifest, supersymmetry which is non-linearly realized and has the form 
\begin{equation}
\delta^* W_\alpha = 2 m \, \eta_\alpha 
+\frac12 \overline D^2 \overline{\cal X} \, \eta_\alpha 
+ 2i  \partial_{\alpha\dot\alpha} {\cal X} \, \overline \eta^{\dot\alpha} \, ,
\quad 
\delta^* {\cal X} =  \eta^\alpha W_\alpha \, ,
\label{BGNLsusy} 
\end{equation}
where $\eta_\alpha$ denotes the global parameter associated to this second, non-linear, supersymmetry. The component form of \eqref{BG1} reads 
\begin{equation}
{\cal L}_\text{BG} = \beta m \left(1-\sqrt{1+\frac{1}{2m^2}(F^2-2 \text{D}^2)-\frac1{16m^4}(F\tilde F)^2}\right) + \text{fermions} \, ,  
\end{equation}
where $\tilde F_{mn}=\frac12\varepsilon_{mnkl} F^{kl}$ and, once the auxiliary field D is integrated out, the Lagrangian reduces to
\begin{equation} \label{BGwithoutF}
{\cal L}_\text{BG} = \beta m\left(1 - \sqrt{ - \det \left(n_{mn} + \frac 1mF_{mn}\right) } \right) + \text{fermions} \, .  
\end{equation}

First we would like to recast this action into an equivalent one, in which the spontaneously broken non-linear supersymmetry becomes manifest and is described by a Volkov--Akulov fermion. This would allow us to take into account all the fermionic contributions and to deal with them in a manifestly supersymmetric way, once the Lagrangian is lifted to superspace. A result similar to this has been obtained in \cite{Bellucci:2015qpa} with the use of the coset construction for implementing the non-linear realization of supersymmetry. 

Following \cite{Klein:2002vu}, we define the fermionic superfield 
\begin{equation}
\Gamma^W_\alpha =  \frac{W_\alpha}{2 m + \frac12 \overline D^2 \overline {\cal X}} \, 
\end{equation} 
which, in our conventions, has non-canonical mass dimension -1/2. From the non-linear supersymmetry transformations \eqref{BGNLsusy}, we can see that its lowest component $\Gamma^W_\alpha | = \gamma^W_\alpha$, transforms as 
\begin{equation}
\label{dgammaW}
\delta^* \gamma^W_\alpha = \eta_\alpha 
- 2i  \gamma^W \sigma^m \overline \eta \ \partial_m \gamma^W_\alpha \, ,
\end{equation}
which is precisely the supersymmetry transformation of the alternative goldstino spinor $\gamma_\alpha$ discussed in the Appendix \ref{AGS}. The fermion $\gamma_\alpha^W$ is a function of the component fields of the vector multiplet, namely of $v_m$, $\chi_\alpha$ and D. As a consequence, it is always possible to interchange $\chi_\alpha$ with $\gamma^W_\alpha$ by means of a field redefinition. It is now important to observe that the Lagrangian \eqref{BG1}, 
up to an additive constant ($-2 \beta m$) and boundary terms, has the form 
\begin{equation}
\label{BG2}
{\cal L}_\text{BG} = {\cal B} + \overline{\cal B} \, , 
\quad 
{\cal B} = -\beta \left( m + \frac14 \overline{D}^2 \overline{\cal X}|\right) \,  
\end{equation}
and transforms under \eqref{BGNLsusy} in a specific way 
\begin{equation}
\label{dBG} 
\delta^* {\cal L}_\text{BG} = 
-2 i \, \partial_a \left( \gamma^W \sigma^a \overline \eta \, {\cal B} \right)
-2i \, \partial_a \left( \overline \gamma^W \overline \sigma^a \eta \, \overline{\cal B} \right) \, . 
\end{equation} 
As we prove in the Appendix \ref{AGS}, any Lagrangian of the form \eqref{BG2}, which transforms under supersymmetry as \eqref{dBG}, is equivalent to a Lagrangian of the type \eqref{SNLL} written in terms of the standard non-linear realizations of supersymmetry, with the goldstino transforming as the Volkov--Akulov fermion \eqref{gold-S3}. The Bagger--Galperin action can therefore be recast into the equivalent form 
\begin{equation}
\label{SBG2}
S_\text{BG} = \int d^4 x \, \det[A_m^{\ a}] \, \hat{ L}_\text{BG} \, , 
\end{equation}
where the field $\hat{ L}_\text{BG}$ is uniquely fixed by the non-linear dressing with the Volkov--Akulov goldstino $\lambda_\alpha$, 
namely the operation $e^{\delta^*_\eta} (\ldots) |_{\eta=-\lambda}$, which is presented in the Appendix \ref{Appendix3}:\footnote{As a consequence of the uniqueness of the non-linear dressing that we prove in Appendix \ref{Appendix1}, when lifted to superspace this particular dressing is actually equivalent to the one introduced in Appendix \ref{Appendix1} and performed at the superfield level.}  
\begin{equation}
\label{LBG33}
\begin{aligned}
 \hat{ L}_\text{BG} = & e^{\delta^*_\eta} \, \left( {\cal B} + \overline{\cal B} \right) |_{\eta=-\lambda}
\\[.1cm] 
= & e^{\delta^*_\eta} \, \left(- \beta m -\beta m\sqrt{1+\frac{1}{2m^2}(F^2-2\text{D}^2)-\frac1{16 m^4}(F\tilde F)^2} \right) \Big{|}_{\eta=-\lambda} \, . 
\end{aligned} 
\end{equation} 
In particular, in this new description the Volkov--Akulov goldstino $\lambda_\alpha$ is a function of the gaugino $\chi_\alpha$, via its dependence on $\gamma^W_\alpha$, which is given by \cite{Samuel:1983jp} 
\begin{equation}
\label{VAlambda}
\begin{aligned}
\lambda_\alpha = \gamma^W_\alpha 
& - i t^a \partial_a  \gamma^W_\alpha 
- \frac12 t^a t^b \partial_a \partial_b \gamma^W_\alpha  
- t^a \partial_a t^b \partial_b \gamma^W_\alpha 
\\
& + i t^a \partial_a t^b \partial_b t^c \partial_c \gamma^W_\alpha 
+ i t^a t^b \p_a \p_b t^c \p_c \gamma^W_\alpha \, , 
\end{aligned}
\end{equation}
where $t_a = \gamma^W \sigma_a \overline \gamma^W$. As a consequence, the gaugino satisfies 
\begin{equation}
\lambda^2 \overline \lambda^2 \chi_\alpha =0 \, , 
\end{equation}
which in turn, due to the property \eqref{prop2} of the dressing operators, leads to
\begin{equation}
e^{\delta^*_\eta} \, \chi|_{\eta=-\lambda} =   e^{\delta^*_\eta} \, \lambda(\ldots)|_{\eta=-\lambda} = 0 \, . 
\end{equation}
The same property is used to derive the second line in \eqref{LBG33} from the first line. Notice that we have been able to pass from the standard Bagger-Galperin action \eqref{BG1} to \eqref{SBG2}, while keeping explicitly track of all the fermionic contributions. This can be achieved since, in the original Bagger-Galperin model, there is only one fermion, which has the role of the goldstino associated to the partial breaking of supersymmetry and which is used to implement the non-linear realization. 
The action of the dressing operators on a collection of fields is the same as their action on all the fields individually, 
therefore the Lagrangian in \eqref{SBG2} can be equivalently written as 
\begin{equation}
\begin{aligned}
\label{BGinterm}
{\cal L}_\text{BG} =&  -\beta m   \det[A_m^a] \times 
\\
&\left( 1+\sqrt{1+\frac{1}{2m^2}[(e^{\delta^*_\eta}F^2)|_{\eta=-\lambda}-2 (e^{\delta^*_\eta}\text{D}^2)|_{\eta=-\lambda}]-\frac1{16m^4}e^{\delta^*_\eta}(F\tilde F)^2|_{\eta=-\lambda}} \right) . 
\end{aligned}
\end{equation}

The next step is to carry out explicitly the field redefinitions inside the square root, namely the dressing with the goldstino $\lambda_\alpha$, in order to produce all the necessary fermionic terms which are implementing the non-linear realization of the spontaneously broken supersymmetry. After this procedure, the redefined fields will transform as a standard non-linear realization of supersymmetry. While the case of the scalar auxiliary field is straightforward, that of the vector requires some care, due to the associated gauge invariance. From \eqref{BGNLsusy} we can deduce the variation of the vector $v_a$ under the second supersymmetry which, because of gauge invariance, is defined only up to a gauge transformation. We have indeed
\begin{equation}
\label{v2susy}
\delta^* v_a = - G \sigma_a \overline \eta - \eta \sigma_a \overline G  - i  \, \partial_a( \lambda \sigma^b  v_b ) \overline \eta + i  \, \eta  \partial_a( \sigma^b v_b \overline \lambda) \, , 
\end{equation}
where $G_\alpha$ is a function of the components of the vector multiplet, defined as 
\begin{equation}
\begin{aligned}
G_\alpha &= D_\alpha {\cal X}| \\
&= -\frac{1}{2m} W^\beta D_\alpha W_\beta|-\frac{1}{8m}D_\alpha \overline D^2 \left(\frac{W^2 \overline W^2}{4 m^2 + \frac12 A_+ + \sqrt{16 m^4+4m^2 A_+ + \frac12 A_-^2}}\right)\bigg |\\
&=-\frac{1}{2m}\left(i \chi_\alpha \text{D} - \sigma^{cd\, \beta}_{\ \ \alpha} \chi_\beta F_{cd} \right)\left[\frac{-4 \text{D}^2+ 2 F_{ab}F^{ab}-2i F_{ab}\tilde F^{ab}}{4 m^2 + \frac12 a_+ + \sqrt{16 m^4 + 4 m^2 a_+ + \frac12 a_-^2}}\right]+{\cal O}(\chi^2),
\end{aligned}
\end{equation}
with $a_+ = A_+|$ and $a_-= A_-|$. We recall that the Volkov--Akulov goldstino $\lambda_\alpha$ is also a function of the vector superfield components, given by \eqref{VAlambda}. The last two terms in \eqref{v2susy} are clearly just a gauge transformation and we are allowed to choose them in order that in the field redefinition
\begin{equation}
u_m=A_m^a \left(e^{\delta^*_\eta}v_a\right)|_{\eta=-\lambda}=v_m+{\cal O}(\lambda^2)\label{redvec}
\end{equation} 
all terms of linear order in the goldstino vanish. One can check then that the redefined vector in \eqref{redvec} transforms indeed in the standard non-linear way
\begin{equation}
\delta^*_\eta u_m=-\rmi\left(\lambda\sigma^n\overline\eta
-\eta\sigma^n\overline\lambda\right)\partial_n u_m-\rmi\partial_m\left(\lambda\sigma^n\overline\eta
-\eta\sigma^n\overline\lambda\right)u_n.
\end{equation}
Due to \eqref{redvec} the field strength of $u_m$ differs from that of $v_m$ appearing in \eqref{BGinterm} by terms with at least one bare goldstino, i.e. a goldstino which is not appearing inside a derivative. The terms with bare $\lambda$ arising from the redefinition \eqref{redvec} will vanish by the properties of the dressing operators given in Appendix \ref{Appendix3}. We have 
\begin{align}   
e^{\delta^*_\eta}F_{ab}|_{\eta=-\lambda}&= e^{\delta^*_\eta}(\partial_{a} v_{b}-\partial_{b} v_{a})|_{\eta=-\lambda}\nonumber\\
&=e^{\delta^*_\eta} ( \delta_a^m \delta_b^n (\partial_{m} u_{n}-\partial_{n} u_{m}))|_{\eta=-\lambda}\nonumber\\
&=e^{\delta^*_\eta}((A^{-1})_a^m(A^{-1})_b^n\left[\partial_{m} u_{n}-\partial_{n} u_{m}\right])|_{\eta=-\lambda}\,,
\end{align} 
where the term that appears on the last line inside the brackets is the definition of a field strength in the standard non-linear realization. Let us also present for completeness the supersymmetry transformation of D, which reads 
\begin{equation}
\label{DDD}
\delta^*_\eta \text{D} = i \, \eta^\alpha \partial_{\alpha \dot \alpha} \overline G^{\dot \alpha} 
+ i \, \overline \eta^{\dot \alpha} \partial_{\alpha \dot \alpha} G^{\alpha} \, .  
\end{equation}
To sum up, the uniqueness theorem of the dressing operators together with their properties proved in the Appendix \ref{Appendix3}, allow us to replace all field strengths in the action \eqref{BGinterm} with their standard non-linear counterparts. 
We obtain 
\begin{equation} 
\label{space}
S_\text{BG}=\beta m\int d^4 x \det[A_m^a]
\left(-1 -\sqrt{1+\frac{1}{2m^2}({\cal F}^2-2 {\cal D}^2)-\frac1{16m^4}({\cal F}\tilde {\cal F})^2} \right) \, , 
\end{equation}
where
\begin{equation}
{\cal F}_{ab}=(A^{-1})^m_a(A^{-1})^n_b\left[\partial_{m} u_{n}-\partial_{n} u_{m}\right] 
\end{equation}
and where we have redefined the scalar D as follows 
\begin{equation} 
{\cal D}=e^{\delta^*_\eta} \text{D}|_{\eta=-\lambda}\,. 
\end{equation} 
The new scalar ${\cal D}$ transforms as \eqref{SNLsf}. It should be considered as a fundamental real scalar field and not as composite anymore, because we are performing a field redefinition to pass from D to ${\cal D}$. This does not apply to ${\cal F}_{ab}$, which is not just the field strength of the redefined vector $u_m$, but it contains goldstino interactions as well. Finally when we integrate out the auxiliary field ${\cal D}$, the action reduces to the Born--Infeld, up to the additive constant we introduced in the start
\begin{equation} 
\label{LBGF}
S_\text{BG}=- \beta m \int d^4 x \det[A_m^a] \left( 
1+ 
\sqrt{-\det{\left(\eta_{ab}+\frac{1}{m}{\cal F}_{ab}\right) }} 
\right) \, .  
\end{equation} 
This is a component form of the Bagger--Galperin action in which the non-linear realization of supersymmetry 
is manifest \cite{Bellucci:2015qpa}. 
In the next subsections we are going to lift this action to superspace and discuss its relationship with the new D-term introduced in \cite{Cribiori:2017laj}. 
Let us mention that if we embody the Born--Infled action into a string theory setup, the scales $\sqrt m$ and $\sqrt \beta$ are related to the $\alpha'$ and to the brane tension T as follows 
\begin{equation}
\beta m = \text{T} \,  , \quad\frac{1}{2 \pi \alpha'} = m \, , 
\end{equation}
while the gauge coupling is given by $1/g^2 = \beta/m$, which leads to
\begin{equation}
g = \frac{1}{2 \pi \alpha' \sqrt{\text{T}}} \, .
\end{equation}
As a last remark, 
notice that if we truncate the gauge vector by setting ${\cal F}_{ab} = 0,$ which gives $u_m=\partial_m \phi$ (we recall that $\phi$ transforms under supersymmetry as \eqref{SNLsf}), the action \eqref{LBGF} reduces to the Volkov--Akulov \cite{Volkov:1973ix}.

\subsection{The alternative Bagger--Galperin action}
\label{sec_splitt}

In the previous subsection we have shown that the Bagger-Galperin action can assume the component form \eqref{LBGF}, in which supersymmetry is manifestly non-linearly realized and the goldstino is the Volkov--Akulov fermion $\lambda_\alpha$. We stress however that, on top of the manifest and spontaneously broken supersymmetry, a second, unbroken supersymmetry is present in the model from the very beginning, as proved in \cite{Bellucci:2015qpa}, even if this might not be obvious from the form of the action \eqref{LBGF}. 

In this subsection we embed the action $\eqref{LBGF}$ into $\mathcal{N}=1$ superspace. We are going to give two different superspace descriptions and then we show their equivalence. One of these two descriptions gives rise to the new D-term proposed in \cite{Cribiori:2017laj}. Even though in global supersymmetry these alternative formulations are equivalent to the original Bagger--Galperin model, their coupling to supergravity will differ from the latter. This is in accordance with the interpretation of the original Bagger--Galperin action coupled to supergravity as an effective description of a probe D3-brane in a curved background, while the alternative formulations as that of a probe anti D3-brane. 

As discussed extensively in the Appendix \ref{appA}, the superspace embedding of an action of the type $\eqref{LBGF}$ is given by applying formula \eqref{ALDL} and takes the form 
\begin{equation}
S_{\overline{\text{BG}}}=- \beta m \int d^4x\, d^4\theta\,\Lambda^2\overline\Lambda^2 
\left[1+ \sqrt{-\det\left(\eta_{ab}+ \frac{1}{m} {\mathbb F}_{ab} \right)}
\right]
\,, 
\label{BIalmost} 
\end{equation}
where $\Lambda_\alpha$ is the goldstino superfield satisfying the properties \eqref{SW} and which contains $\lambda_\alpha$ in the lowest component, while $\mathbb{F}_{ab}$ is the dressed field strength of the vector, defined in \eqref{FFAA}. The quantity $\Lambda^2\overline \Lambda^2$ describes the goldstino sector, while the square root contains the couplings to the vector field, which are written in an appropriate way in order to have a Lagrangian invariant under non-linearly realized supersymmetry.
We refer the reader to the Appendix for more details about these objects and their definitions.

The action \eqref{BIalmost} has the spontaneously broken and non-linearly realized supersymmetry manifestly described by superspace, as desired. This is indeed one of the results of the paper and we have denoted it as $S_{\overline{\text{BG}}}$ because, 
when coupled to supergravity, it can be interpreted as the effective action of an anti D3-brane in a curved background. 
In the Volkov--Akulov description of the goldstino sector, however, it is not clear what mechanism sources the breaking of supersymmetry. Moreover, since we would like to relate the action \eqref{BIalmost} to the new D-term of \cite{Cribiori:2017laj}, which is described by means of an $\mathcal{N}=1$ vector multiplet $V$, an alternative superspace formulation of \eqref{LBGF} in terms of $V$ is needed. Following this logic, we propose the superspace action 
\begin{equation}
\begin{aligned} 
S_{\overline{\text{BG}}}=&\frac{\beta}{4m} \int d^4 x\, \left( d^2\theta\, W^2 + \hc \right) 
+ 16  \beta  \int d^4x\, d^4\theta\, \frac{W^2\overline{W}^2}{D^2W^2\overline D^2\overline W^2}D^\alpha W_\alpha\\
& + 16 \beta m \int d^4x\, d^4\theta\, \frac{W^2\overline{W}^2}{D^2W^2\overline D^2\overline W^2}\left\{1+\frac{1}{4m^2} f_{ab} f^{ab} -\sqrt{-\det \left(\eta_{ab} + \frac{1}{m} f_{ab} \right) } \right\} \,,
\label{Dtermgen}
\end{aligned}
\end{equation}
where we have defined the superfield\footnote{We would like to point out that $f_{ab}$ can be also defined as $f_{ab}=  \partial_a V_b - \partial_b V_a$, where $V_a$ is given by \eqref{V_a}, since it is a superfield with lowest component the field strength of the gauge vector, namely $f_{mn}| = F_{mn}$.}
\begin{equation}
f_{ab}=\frac{\rmi}{4}\sigma_{ab\,\gamma}{}^\alpha\varepsilon^{\gamma\beta}\left(D_\alpha W_\beta +D_\beta W_\alpha\right)+\hc\,.
\label{BGNL} 
\end{equation}
Notice that the first line of \eqref{Dtermgen} has the same structure as the new D-term of \cite{Cribiori:2017laj}. 
The rest of this subsection is devoted to show the equivalence between \eqref{Dtermgen} and \eqref{BIalmost}. The reader who is not interested in the proof can skip the present subsection at first reading.

Since the action \eqref{LBGF} does not contain auxiliary fields, as a first step in order to prove the aforementioned equivalence we parametrize the vector $V$ in order to separate its auxiliary degrees of freedom from the propagating ones. 
To this purpose, we split it into two pieces\cite{Cribiori:2017ngp} 
\begin{equation}
V=\check V +\frac12\Phi\overline\Phi {\cal A}\,, \label{splitting}    
\end{equation}
where $\Phi$ is the constrained nilpotent superfield \cite{Rocek:1978nb,Casalbuoni:1988xh} defined in \eqref{rocekde}, namely 
\begin{equation}
\Phi=-\frac14 \overline D^2\left(\Lambda^2\overline\Lambda^2\right)\label{rocek}\,, 
\end{equation}
while $\check V$ and $\mathcal{A}$ are respectively a constrained vector and chiral superfield, as explained in a while. The mass dimension of $\Phi$ is -1, while that of $\mathcal{A}$ is 2, therefore both $V$ and $\check V$ have vanishing mass dimension. The fact that $\mathcal{A}$ has mass dimension 2 is confirming that it represents auxiliary degrees of freedom. Indeed such a superfield contains only one independent degree of freedom in its lowest component, which we identify with D and which is real as a consequence of the constraint
\begin{equation}
\Phi{\cal A}=\Phi\overline{\cal A} \label{constrA}\,. 
\end{equation}
Since the superfield $\Phi$ describes only the goldstino, in order to match the number of degrees of freedom on both sides of \eqref{splitting}, also the vector superfield $\check V$ has to be constrained. We define therefore the field strength superfield $\check W_\alpha=-\frac14\overline D^2 D_\alpha \check V$ and we constrain it as
\be
\begin{aligned}
\Phi \check W_\alpha=0\,,\\
\Phi\overline\Phi D^\alpha\check W_\alpha=0\,.  \label{Wconstr}
\end{aligned}
\ee
The first constraint is removing the fermion from $\check V$, while the second eliminates the real auxiliary fields, since they are already described by $\Phi$ and $\mathcal{A}$ respectively. From these equations and the splitting \eqref{splitting} one can furthermore derive the following important relation 
\begin{equation}
\frac{\Phi}{D^2\Phi}=\frac{W^2}{D^2 W^2} \label{rocekW}\,.
\end{equation}
By using the splitting \eqref{splitting} together with the constraints \eqref{constrA}, 
\eqref{Wconstr} and \eqref{rocekW}, the form of \eqref{Dtermgen} can be simplified. The result is an action in which there are no couplings between the constrained superfields $\check W_\alpha$ and $\mathcal{A}$, which describe respectively the vector and the real auxiliary field. This action takes the form
\begin{equation}
\begin{aligned}
S_{\overline{\text{BG}}}  = & -\frac{\beta}{16 m}\int d^4 x\, d^4\theta\,\Phi\overline\Phi \left[ D^2\check W^2 +\hc \right]+\frac{\beta}{2 m} \int d^4x d^4\theta\,\Phi\overline\Phi{\cal A} \overline{\cal A} \\
& -  \beta  \int d^4x d^4\theta\,\Phi\overline\Phi \left( {\cal A} + \overline{\cal A} \right) \\
& +\frac{\beta m}{8}\int d^4x\, d^4\theta\,\Phi\overline\Phi \left\{8 + \frac{2}{m^2} \check f_{mn} \check f^{mn} -8\sqrt{-\det \left( \eta_{mn}+ \frac{1}{m} \check f_{mn} \right) }
\right\} \,. 
\label{SGenDtermPhiA}
\end{aligned}
\end{equation}
Since the auxiliary field has completely been separated from the rest of the action, it can now be integrated out straightforwardly. 
We first take into account the constraint in \eqref{constrA} by adding a Lagrange multiplier $U$. The field equations of the relevant part of the action 
\begin{equation}
\begin{aligned}
\mathcal{S}_{\overline{\text{BG}},\,\text{aux}}=& \, \frac12 \int d^4x d^4 \theta \, \Phi\overline\Phi\left[ \frac{\beta}{m} {\cal A} \overline{\cal A} 
- 2 \beta  \left({\cal A} + \overline{\cal A}\right)\right]\\
&+\left(\int d^4 x d^4\theta \, U\left(\Phi{\cal A}-\Phi\overline{\cal A}\right)+\hc\right) \,, 
\end{aligned}
\end{equation}
are then for ${\cal A}$ and $U$ respectively 
\begin{align}
\label{EEE1}
\delta {\cal A}\;&:\;-\frac18\overline D^2\left[
\Phi\overline\Phi 
\left( \frac{\beta}{m} \overline{\cal A} - 2 \beta  
\right) 
\right] 
-\frac14\overline D^2\left(U\Phi-\overline U\overline \Phi\right)=0 \, ,  \\
\delta U\;&:\;\Phi{\cal A}-\Phi\overline{\cal A}=0\,.
\end{align} 
We can now multiply the equation \eqref{EEE1} with $\overline \Phi$ and divide by $\overline D^2\overline \Phi$ to obtain
\begin{equation}
\overline \Phi \overline U=-\frac12 \beta\Phi\overline\Phi\left(\frac{\mathcal{A}}{m}-2\right)-\frac14\overline\Phi\overline D^2\left(\Phi U\right)\,.
\end{equation}
By iterating for $\overline \Phi \overline U$
\begin{align}
\overline \Phi \overline U &=\frac12 \beta\Phi\overline\Phi\left(\frac{\mathcal{A}}{m}-2\right)-\frac14\overline\Phi\overline D^2\left(\Phi U\right)\nonumber\\
&=\beta\Phi\overline\Phi\left(\frac{\mathcal{A}}{m}-2\right)-\frac14\Phi D^2\left(\overline\Phi\overline U\right)\nonumber\\
&= \frac32 \beta\Phi\overline\Phi\left(\frac{\mathcal{A}}{m}- 2\right)-\frac14\overline\Phi\overline D^2\left(\Phi U\right)  
\end{align}
and comparing the first line with the last, we finally get
\begin{equation}
\Phi\overline\Phi({\cal A} - 2 m )=0 \, .
\end{equation}
Once this result, which is a consequence of the equations of motion, is implemented inside the action \eqref{SGenDtermPhiA}, 
the terms containing $\mathcal{A}$ add with the first term in the third line giving 
\begin{equation} 
\frac{\beta}{2m}\int d^4x d^4\theta \Phi \overline \Phi \mathcal{A}\overline{\mathcal{A}}-\beta\int d^4xd^4\theta \Phi \bar \Phi (\mathcal{A}+\overline{\mathcal{A}})+\beta m \int d^4 x d^4\theta \Phi\overline {\Phi}=-\beta m\int d^4x\, d^4\theta\,\Phi\overline\Phi \, . 
\end{equation}
Then, by using that 
\begin{equation}
\Phi \overline \Phi (D^2 \check W^2 +c.c.) = 4 \Phi \overline \Phi \check f_{mn}\check f^{mn} \, , 
\end{equation}
the first part in \eqref{SGenDtermPhiA} cancels the second term in the third line and all that is left is the determinant term supported by the goldstino superfields $\Phi\overline\Phi$, namely 
\begin{equation}
\label{PhiBG}
S_{\overline{\text{BG}}}=- \beta m \int d^4x\, d^4\theta\,\Phi\overline\Phi 
\left[ 1+ \sqrt{-\det\left(\eta_{mn}+ \frac{1}{m} \check f_{mn}\right)} 
\right] 
\, . 
\end{equation} 
Notice that the Lagrangian \eqref{PhiBG} is written entirely in terms of constrained superfields, i.e. the nilpotent goldstino superfield $\Phi$ and the constrained gauge vector superfield $\check V$. We have therefore provided a formulation of the effective theory of the anti D3-brane within the constrained superfields approach of non-linear supersymmetry. By using \eqref{phiL2}, we can recast \eqref{PhiBG} in the form 
\begin{equation}
\label{LSN5}
S_{\overline{\text{BG}}}=- \beta m \int d^4x\, d^4\theta\, \Lambda^2 \overline \Lambda^2 
\left[1+ \sqrt{-\det\left(\eta_{mn}+ \frac{1}{m} \check f_{mn}\right)} 
\right] 
\, . 
\end{equation}

We have shown that the original action \eqref{Dtermgen} is on-shell equivalent to \eqref{LSN5}. It remains now to relate \eqref{LSN5} to \eqref{LBGF}, in order to conclude the proof and demonstrate that \eqref{Dtermgen} is indeed an alternative form of the Bagger--Galperin action. To this end we dress the vector and its field strength with goldstino interactions, in such a way that the resulting expressions transform as a  standard non-linear realization of supersymmetry.  The necessary ingredients are discussed thoroughly in the Appendix \ref{appA}. In particular, we embed the vector $v_m$ into the superfield $\check V_a$ defined in \eqref{V_a} and we perform a field redefinition of the form 
\begin{equation}
{\mathbb V}_m = \frac{1}{16} \, {\mathbb A}_m^a \, \Pi^2 \overline \Pi^2 \left( \Lambda^2 \overline \Lambda^2 \, \check V_a \right) , 
\end{equation} 
where $\Lambda_\alpha$ is the goldstino superfield \eqref{SW}, $\mathbb{A}_m^a$ is the (superfield) vielbein \eqref{Aam}, which depends on the goldstino, and $\Pi_\alpha$ is the superspace derivative \eqref{Pdef}. Out of the new vector superfield $\mathbb{V}_m$, which is a function of the original vector superfield $V$ and of the goldstino, we can then construct the field strength
\begin{equation}
{\mathbb F}_{ab} = ({\mathbb A}^{-1})_a^m ({\mathbb A}^{-1})_b^n \left( \p_m {\mathbb V}_n - \p_n {\mathbb V}_m \right).
\end{equation}
This field strength $\mathbb{F}_{ab}$ is an object of the form \eqref{Tprop}, with $T$ replaced by the superfield $\check f_{ab}$. As a consequence it holds that
\begin{equation}
\Lambda^2 \overline \Lambda^2 {\mathbb F}_{ab} = \Lambda^2 \overline \Lambda^2 \check f_{ab} \, 
\end{equation}
and the action \eqref{LSN5} then reduces to \eqref{BIalmost}. We have thus concluded the demonstration proving that the Bagger--Galperin action \eqref{BG1} is equivalent to the alternative form \eqref{Dtermgen}. In the following we will study some properties of \eqref{Dtermgen} and we will couple the system to supergravity.

\section{Bagger--Galperin action and D-term breaking models} 

After having elaborated on the possible forms the Bagger--Galperin action can assume, we would now like to concentrate on its relationship with the different D-term supersymmetry breaking mechanisms which have been proposed in the literature \cite{Fayet:1974jb, Cribiori:2017laj}. First of all, we will show that it is possible to obtain the new D-term breaking of \cite{Cribiori:2017laj} by taking an appropriate limit of the Bagger--Galperin model. In this sense, the new D-term is incorporated in the Bagger--Galperin model without the need for additional modifications. On the contrary, as we will show, the structure of the standard Fayet--Iliopoulos term differs significantly from that of the Bagger--Galperin action. Finally it is known that it is possible to add to the Bagger--Galperin Lagrangian a contribution of the type of the standard Fayet--Iliopoulos D-term \cite{Ivanov:1997mt, Antoniadis:2008uk,Kuzenko:2009ym}. We will therefore discuss briefly the differences with these constructions and leave details for the Appendix \ref{appB}.

\subsection{Weak-field expansion of alternative Bagger--Galperin} 

In this subsection we analyze the relationship between the Bagger--Galperin model and the global limit of the new Fayet--Iliopoulos D-term of \cite{Cribiori:2017laj}, namely 
\begin{equation}
S_\text{New FI} =  \frac{1}{4g^2} \int d^4 x\, \left[ d^2\theta\, W^2 + \hc \right] 
+ 16 \xi \int d^4x\, d^4\theta\, \frac{W^2\overline{W}^2}{D^2W^2\overline D^2\overline W^2}D^\alpha W_\alpha \, ,
\label{newDtermglobal}
\end{equation} 
whose vacuum energy is given by ${\cal V} =  2 g^2 \xi^2$. For convenience we will compare the Lagrangian \eqref{newDtermglobal} to the form \eqref{Dtermgen} of the Bagger--Galperin action, which we have shown to be equivalent to \eqref{BG1}. 

The starting point is to notice that the first line of \eqref{Dtermgen} matches with the full expression \eqref{newDtermglobal} of the new D-term in the global limit, once we relate the parameters in the following way 
\begin{equation}
\xi =  \beta = 2  \pi \alpha' \text{T} \, , \qquad g = \sqrt{m / \beta} = (2 \pi \alpha' \sqrt{\text{T}})^{-1}. 
\end{equation} 
We concentrate now on the second line of \eqref{Dtermgen}, in order to understand if there exists a physical regime in which it can be consistently ignored, when compared to the first. The terms of interest are
\begin{equation}
\label{ST1}
16 \beta m \int d^4x\, d^4\theta\, \frac{W^2\overline{W}^2}{D^2W^2\overline D^2\overline W^2}\left\{1+\frac{1}{4m^2} f_{ab} f^{ab} -\sqrt{-\det \left(\eta_{ab} + \frac{1}{m} f_{ab} \right) } \right\},
\end{equation}
where we recall that $f_{ab}$ is the superfield defined in \eqref{BGNL}. By using the results presented in the Appendix \ref{appA} and following a reasoning similar to that of the previous section, 
we see that \eqref{ST1} takes the form  
\begin{equation}
\begin{aligned}
& \beta m \int d^4x\, d^4\theta\, \Phi \overline{\Phi}  
\left\{1+\frac{1}{4m^2} \check f_{ab} \check f^{ab} -\sqrt{-\det \left(\eta_{ab} + \frac{1}{m} \check f_{ab} \right) } \right\}
\\
&= \beta m \int d^4x\, d^4\theta\, \Lambda^2 \overline \Lambda^2 \,
\left\{ 1+\frac{1}{4 m} {\mathbb F}_{mn} {\mathbb F}^{mn} - \sqrt{-\det \left( \eta_{mn} + \frac{1}{m} {\mathbb F}_{mn} \right)}\right\}
\\
&= \beta m \, \int d^4x\, \det[A_m^a] \, \left\{ 1+\frac{1}{4 m} {\cal F}^2 - \sqrt{-\det\left(\eta_{mn} + \frac1m {\cal F}_{mn}\right)}
\right\} \, ,
\label{ST3}
\end{aligned}
\end{equation}
where we have parametrized the vector superfield according to \eqref{splitting}. In particular, to pass from \eqref{ST1} to \eqref{ST3} we used that $\Phi \overline \Phi f_{ab} = \Phi \overline \Phi \check f_{ab}$, while to pass from the second to the third line in \eqref{ST3} we employed the identity \eqref{ALDL}. If we expand then the square root of the determinant we have
\begin{equation}
\label{EXP}
\sqrt{-\det \left( \eta_{mn} + \frac 1m {\cal F}_{mn} \right) } =  1 + \frac{1}{4m^2} {\cal F}^2 + {\cal O}({\cal F}^4) \,  
\end{equation}
which, once inserted into \eqref{ST3}, reveals that only ${\cal O}({\cal F}^4)$ terms contribute to the second line in \eqref{Dtermgen}. 
We have found therefore that the Bagger--Galperin Lagrangian and the new D-term are related as
\begin{equation}
\label{BGv=Dterm}
{\cal L}_{\overline{\text{BG}}} = {\cal L}_\text{New FI} + {\cal O}({\cal F}^4) \, , 
\end{equation}
where the Lagrangian in \eqref{newDtermglobal}  takes the form 
\be
{\cal L}_\text{New FI} = - 2 \xi^2 g^2 \det[A_p^e] - \frac{1}{4 g^2} \det[A_p^e] {\cal F}_{mn}{\cal F}^{mn} \, . 
\ee
For effective theories in which the electro-magnetic field $F_{mn}$ is weak, the terms ${\cal O}({\cal F}^4)$ can be ignored. We then conclude that the new Fayet--Iliopoulos D-term arises in the weak-field limit of the Bagger--Galperin action.\footnote{Different constructions can be found in \cite{Kuzenko:2017zla,Kuzenko:2018jlz} and it would be interesting to see if and how they relate to D-branes.}

\subsection{Standard Fayet--Iliopoulos term}

We would like now to compare and contrast the weak-field limit of the Bagger--Galperin action with the original D-term proposed by Fayet and Iliopoulos in \cite{Fayet:1974jb}, 
namely 
\begin{equation}
\label{SFI}
{\cal L}_\text{standard FI} = \frac14 \left(\int d^2\theta\, W^2 +c.c.\right) 
- 2 \sqrt 2 \, \tilde \xi \int d^4 \theta\, V \, . 
\end{equation}
As we are going to argue, a mismatch is present between the two models, which points to the fact that not all theories with a D-term breaking will match with the Bagger--Galperin in the weak-field limit. Within the framework of non-linear supersymmetry, the Lagrangian \eqref{SFI} has been recast in the languange of constrained superfields in \cite{Cribiori:2017ngp}. Its form is
\begin{equation}
\label{LLXV}
{\cal L}_\text{standard FI} = - \int d^4 \theta \, \Phi \overline \Phi + \frac14 \left(\int d^2\theta\, \check W^2 +c.c.\right) 
- 2 \sqrt 2 \, \tilde \xi \int d^4 \theta\, \check V \,  , 
\end{equation} 
where the first superspace integral contains the kinetic term of the goldstino, the second that of the vector field, while the third one encodes additional non-linear interactions, namely 
\begin{equation} 
\label{mmFI0}
\int d^4 \theta\, \check V  = - i \, \epsilon^{klnm} \, \p_k  \lambda \,  \sigma_l \, \partial_n \overline \lambda \, v_m \, + \ldots ,
\end{equation}  
with dots standing for terms of higher order in $\lambda_\alpha$. We notice that, with respect to the analogous result presented in \cite{Cribiori:2017ngp}, in the Lagrangian \eqref{LLXV} the auxiliary field has been eliminated by means of the constraints on $\Phi$. We can now employ the formalism developed in the Appendix \ref{appA} to handle non-linear realizations of supersymmetry and we can recast the model \eqref{LLXV} into the form
\begin{equation}
\begin{aligned}
\label{LLXVNL}
{\cal L}_\text{standard FI} = & - \tilde \xi^2 \det[A_p^e] - \frac14 \det[A_p^e] {\cal F}_{mn}{\cal F}^{mn} \\
& + 2 \sqrt 2 i \, \tilde \xi \, \det[A_p^e] \, \epsilon^{abcd} \, [(A^{-1})_a^{\ n} \, \p_n  \lambda] \, \sigma_b \, [(A^{-1})_c^{\ k} \, \p_k \overline \lambda] \, (A^{-1})_d^{\ m} \, u_m  \, , 
\end{aligned}
\end{equation} 
which is a function of the goldstino $\lambda_\alpha$ and of the dressed vector 
field $u_m = A_m^a \left(e^{\delta^*_\eta}v_a\right)|_{\eta=-\lambda}$.  
If we compare then \eqref{LLXV} with the weak-field limit of the Bagger--Galperin Lagrangian studied previously, see e.g. \eqref{BGv=Dterm}, we can realize that a mismatch is present due to the term
\footnote{To derive \eqref{mmFI} we notice that $\int d^4 \theta\, \check V$ can take the form $\int d^4 \theta\, \Gamma^2 D^2 (\overline \Gamma^2 \overline D^2 \check V) + c.c.$ which, after acting with the $D$ derivatives inside the parentheses, gives $\int d^4 \theta\, ( -8 i \Gamma^2 D^\alpha \overline \Gamma^2 \partial_{\alpha \dot \alpha} \overline D^{\dot \alpha} \check V + \Gamma^2 \overline \Gamma^2 D^2 \overline D^2 \check V) + c.c.$. Using the properties of $\check V$, we find that the latter expression becomes $\int d^4 \theta\, \Lambda^2 \overline \Lambda^2 ( (- \partial_n \Lambda \sigma^k \overline \sigma^n \sigma^m \partial_m \overline \Lambda \check V_k + c.c.) + 2 \partial_m \overline \Lambda \overline \sigma^n \partial^m \Lambda \check V_n )$. From this form one can derive the right hand side of \eqref{mmFI}, by dressing the fields under the non-linear realization, as we have 
described earlier.} 
\begin{equation}
\label{mmFI} 
\int d^4 \theta\, \check V \ \longleftrightarrow \ 2 \sqrt 2 i \, \tilde \xi \, \det[A_m^a] \, \epsilon^{abcd} \, [(A^{-1})_a^{\ n} \, \p_n  \lambda] \, 
\sigma_b \, [(A^{-1})_c^{\ k} \, \p_k \overline \lambda] \, (A^{-1})_d^{\ m} \, u_m \, . 
\end{equation}  
This means that the standard Fayet--Iliopoulos D-term is not matching with the  Bagger--Galperin model, not even up to $\mathcal{O}(\mathcal{F}^2)$.

It is interesting to notice furthermore that the term which creates the mismatch between the Fayet--Iliopoulos and the weak-field limit of the Bagger--Galperin, i.e. $\int d^4 \theta\, \check V$, is also responsible for the gauging of the R-symmetry when the standard Fayet--Iliopoulos term is coupled to supergravity. Indeed, when the term \eqref{mmFI} is lifted  to $\mathcal{N}=1$ supergravity, couplings containing derivatives on the goldstino will essentially become terms with the gravitino 
\begin{equation}
(A^{-1})_a^{\ m} \partial_m \lambda^\alpha \to \hat D_a \lambda^\alpha 
= e_{a}^{\ m} D_m \lambda^\alpha - \frac{1}{2 M_P} \psi_a^\alpha + \ldots , 
\end{equation}
where $\hat D_a \lambda^\alpha$ is the supercovariant derivative of the goldstino \cite{Samuel:1982uh}.  In other words, when the original Fayet--Iliopoulos D-term is coupled to supergravity, the term \eqref{mmFI} is embedded in a locally supersymmetric setup and it generates the coupling 
\begin{equation} 
\label{Rsym}
\frac{i}{\sqrt 2} \, e \, \frac{\tilde \xi}{M_P^2} \, \epsilon^{klmn} \, \psi_k \, \sigma_l \, \overline \psi_m \, v_n \, , 
\end{equation}
which signals the R-symmetry gauging.

Another class of models which have been constructed consists in the addition of the standard Fayet--Iliopoulos D-term to the Bagger--Galperin Lagrangian \cite{Antoniadis:2008uk,Kuzenko:2009ym}, namely to consider 
\begin{equation}
\label{BGFI}
{\cal L}_\text{BG + FI} = {\cal L}_\text{BG} - 2 \sqrt 2 \, \tilde \xi \int d^4 \theta\, V   \, .
\end{equation}
The properties of this theory have been analyzed in \cite{Antoniadis:2008uk,Kuzenko:2009ym}, where it is found that partial supersymmetry breaking takes place, even thought the supersymmetry parameters are rotated by the Fayet--Iliopoulos term. In any case, the theory \eqref{BGFI} contains again the terms which are going to implement the gauging of the R-symmetry when the model is coupled to supergravity. More details on this model can be found in the Appendix B.

\section{Coupling to ${\mathcal N}=1$ supergravity}

In this section we couple the alternative version of the Bagger--Galperin action \eqref{Dtermgen} to $\mathcal{N}=1$ supergravity. We do not discuss the coupling of \eqref{BG1} to supergravity, since it has already been studied in previous works, 
see for example \cite{Cecotti:1986gb,Abe:2015nxa,Abe:2018plc}, 
but we remind the reader that, when the original Bagger--Galperin action \eqref{BG1} is coupled to $\mathcal{N}=1$ supergravity, the system generically preserves only the linear supersymmetry, whereas the non-linear supersymmetry is explicitly broken. 
On the contrary, when coupling the alternative action \eqref{Dtermgen} to supergravity, a system is obtained in which the non-linear and spontaneously broken supersymmetry is gauged, while the other becomes explicitly broken.\footnote{If not gauged, global supersymmetry coupled to gravity is explicitly broken because generically $D_m(\omega) \epsilon_\alpha \ne 0$ for a global spinor $\epsilon_\alpha$. 
On the other hand, examples of curved backgrounds where the supersymmetric DBI action shows to be invariant under a partially broken rigid second supersymmetry have been studied in \cite{Kuzenko:2015rfx}.}

We treat $\mathcal{N}=1$ supergravity using the old-minimal superspace formulation, with the conventions of \cite{Wess:1992cp} and in reduced Planck mass units that set $8\pi G =1$. We do not completely review the formalism here, rather we only retrieve the parts that are relevant for our discussion. We present also the analogous formulation in the tensor calculus setup of \cite{FVP} in Appendix C.

Within our setup, the Lagrangian of four-dimensional ${\cal N}=1$ supergravity takes the form
\begin{equation}
{\cal L}_\text{SG} = \int d^2 \Theta \, 2 {\cal E} \left( -3 {\cal R} + W_0 \right) + \hc \, , 
\end{equation}
where $W_0$ is a complex constant, contributing to the gravitino mass and negatively to the vacuum energy. The properties of the chiral superfield ${\cal R}$ and of the chiral density $2 {\cal E}$ can be found in \cite{Wess:1992cp}. Let us recall that the old-minimal supergravity multiplet contains the following fields: the vierbein $e_m^{\ a}$, the gravitino $\psi_a^\alpha$, a complex scalar auxiliary field $M$ and a real vector auxiliary field $b_a$. In a supergravity setup the derivatives $\partial_a$, $D_\alpha$, and $\overline D_{\dot \alpha}$ are promoted to ${\cal D}_a$, ${\cal D}_\alpha$, and $\overline{\cal D}_{\dot \alpha}$ respectively and satisfy the curved superspace algebra which can be found in \cite{Wess:1992cp}. An abelian vector multiplet is described by the real superfield $V$ and we define 
\begin{equation}
W_\alpha = - \frac14 \left({\overline{\cal D}^2 -8{\cal R}}\right) {\cal D}_\alpha V \, , 
\end{equation} 
which is invariant under the gauge transformation $V \to V + i S - i \overline S$. 
The component fields of the superfield $W_\alpha$ are given by 
\begin{equation}
\begin{aligned} 
W_\alpha | = & - i \chi_\alpha \, , 
\\
({\cal D}_\alpha W_\beta + {\cal D}_\beta W_\alpha) | = & -4i (\sigma^{ba} \epsilon)_{\alpha \beta} \hat D_b v_a 
= 2i (\sigma^{ba} \epsilon)_{\alpha \beta} \hat F_{ab} \, , 
\\ 
{\cal D}^\alpha W_\alpha | = & - 2 \text{D} \, .
\end{aligned}
\end{equation}
The supercovariant derivative of the vector is defined as 
\begin{equation}
\hat D_b v_{\alpha \dot \alpha} = e_{b}^{m} \left\{ D_m v_{\alpha \dot \alpha} 
+ i (\psi_{m \alpha} \overline \chi_{\dot \alpha} + \overline \psi_{m \dot \alpha} \chi_\alpha ) 
+ \frac{i}{2} \psi_m v \overline \psi_a \sigma^a_{\alpha \dot \alpha} 
\right\} \, , 
\end{equation}
where $D_m$ is the covariant derivative which includes the spin-connection $\omega_{ma}^{\ \ \ b}(e,\psi)$. The supersymmetry transformation of the gaugino is 
\begin{equation}
\delta \chi_\alpha = -2 i \xi_\alpha \text{D} - 2 (\sigma^{ab} \xi)_\alpha \hat F_{ab} \, . 
\end{equation}
Finally, the kinetic term for the ${\cal N}=1$ vector mutiplet within supergravity has the form  $\int d^2 \Theta \, 2 {\cal E} \, W^2+\hc$.

We now have all the ingredients at our disposal that are needed in order to couple the alternative Bagger--Galperin Lagrangian to supergravity. We first generalize \eqref{Dtermgen} to curved superspace, which gives
\begin{equation}
\begin{aligned} 
{\cal L}_{\overline{\text{BG}}}=&\frac{\beta}{4m}  \left[ \int d^2\theta\, 2 {\cal E} \, W^2 + \hc \right] 
+ 16 \beta  \int d^4\theta \, E \, \frac{W^2\overline{W}^2}{{\cal D}^2 W^2 \overline{\cal D}^2 \overline W^2} {\cal D}W 
\\
& + 16 \beta m  \int d^4\theta\, E \, \frac{W^2\overline{W}^2}{{\cal D}^2 W^2 \overline{\cal D}^2 \overline W^2} \left\{1+\frac{1}{4m^2} f_{ab} f^{ab}-\sqrt{-\det \left(\eta_{ab} + \frac{1}{m} f_{ab} \right) } \right\} \,,\label{DtermgenSG}
\end{aligned}
\end{equation}
where
\begin{equation}
f_{cd}=\frac{\rmi}{4}\sigma_{cd\gamma}{}^\alpha\varepsilon^{\gamma\beta}\left( {\cal D}_\alpha W_\beta + {\cal D}_\beta W_\alpha\right)+\hc\,.\label{BGNLSG} 
\end{equation}
The total Lagrangian that we consider is then obtained by adding the supergravity sector to \eqref{DtermgenSG}, namely 
\begin{equation}
\label{totSG}
{\cal L}_\text{TOT} = {\cal L}_\text{SG} + {\cal L}_{\overline{\text{BG}}} \, . 
\end{equation}
In order to uncover the physical content of \eqref{totSG}, we would like to rewrite it in component form. 
However, such a task is non-trivial, as highly non-linear expressions are involved. A simplification occurs if we make use of a posteriori information, i.e. of the fact that in \eqref{totSG} supersymmetry is spontaneously broken with
\begin{equation}
\langle \text{D} \rangle \ne 0 \, . 
\end{equation}
As a consequence we are allowed to perform the calculation in an appropriate gauge in which most of the non-linear interactions are not present. Indeed we can write the Lagrangian in the unitary gauge in which the goldstino is set to zero 
\begin{equation}
\text{Unitary Gauge} \ : \ \chi_\alpha = 0 \, . 
\end{equation} 
In this gauge the component form of \eqref{totSG} simplifies considerably and in fact the equations of motion of the auxiliary fields are 
\begin{equation}
\text{D} = 2 m  \,  , \quad M = -3 W_0  \,  , \quad b_a = 0 \, . 
\end{equation}
The on-shell Lagrangian has then the form 
\begin{equation}
\begin{aligned}
e^{-1} {\cal L}_\text{TOT} \Big{|}_{\chi=0} = & - \frac12 R(e, \omega) + 3 |W_0|^2 + \frac{1}{2} \epsilon^{klmn} (\overline \psi_k \overline \sigma_l D_m \psi_n - \psi_k \sigma_l D_m \overline \psi_n) 
\\
- & 
W_0 \, \overline \psi_a \overline \sigma^{ab} \overline \psi_b -  \overline W_0 \, \psi_a  \sigma^{ab}  \psi_b 
- \beta m \left( 1+  
\sqrt{-\det \left(\eta_{ab} + \frac{1}{m} e_a^m e_b^n F_{mn}\right) } 
\right) . 
\end{aligned}
\end{equation}

In case one is interested in the goldstino interactions, 
the supergravity theory \eqref{totSG} can be brought in a form that will allow us to systematically expand the fermionic sector up to second order. This can be achieved by following the procedure we presented in subsection \ref{sec_splitt} and by rephrasing it within a supergravity setup. Since the generalization to supergravity does not present particular complications, we will not reproduce the complete procedure here, rather we will only highlight the relevant steps. We first lift the goldstino spinor superfield $\Gamma_\alpha$, presented in \eqref{ALT}, to local supersymmetry. In this case the superfield satisfies 
\begin{equation}
\begin{aligned}
\label{ALTSG}
{\cal D}_\alpha \Gamma_\beta & =  \epsilon_{\beta \alpha} \left( 1 - 2 \, \Gamma^2 {\cal R} \right)   \, , 
\\[.1cm]
\overline{\cal D}^{\dot \beta} \Gamma^\alpha &= 2 i \, \left( \overline \s^a \, \Gamma \right)^{\dot \beta} \, {\cal D}_a \Gamma^\alpha 
+ \frac12 \, \Gamma^2 {\cal G}^{\dot \beta \alpha} \, , 
\end{aligned}
\end{equation} 
where ${\cal G}_a$ is a superfield of the supergravity sector, which can be found in \cite{Wess:1992cp}. The second step is to perform the splitting of the vector superfield, which has again the form 
\begin{equation}
V=\check V +\frac12 \Phi \overline \Phi {\cal A} \, \label{splittingSG}. 
\end{equation} 
The superfield $\Phi$ appearing in \eqref{splittingSG} is defined as 
\begin{equation}
\Phi = - \frac14 \left({\overline{\cal D}^2 -8{\cal R}}\right) \Gamma^2 \overline \Gamma^2 \, . 
\end{equation}
It satisfies \cite{Lindstrom:1979kq} 
\begin{equation}
\Phi^2=0 \, , \quad \Phi (\overline{\mathcal{D}}^2-8\mathcal{R})\overline\Phi = - 4 \Phi \, , 
\quad \Phi \overline \Phi = \Gamma^2 \overline \Gamma^2 \, , 
\end{equation}
which are the generalization of \eqref{ROCEK} and \eqref{phiL2}. 
In order for the degrees of freedom to match on both sides of \eqref{splittingSG}, the following additional constraints are imposed
\begin{equation}
\label{CSG}
\Phi {\cal A} = \Phi \overline{\cal A} \,  , \quad\Phi \, \check W_\alpha = 0 \,  , \quad\Phi \overline \Phi \, {\cal D}^\alpha \check W_\alpha = 0 \, , 
\end{equation}
where we have defined $\check W_\alpha = - \frac14 \left({\overline{\cal D}^2 -8{\cal R}}\right) {\cal D}_\alpha \check V $. As we have explained in the second section, by performing the splitting, namely formula \eqref{splittingSG}, we want to describe each of the component fields of the vector superfield $V$ with an individual constrained superfield. The third step is to introduce such a splitting inside the Lagrangian \eqref{totSG} and to integrate out the superfield ${\cal A}$ by solving its own equations of motion. This procedure gives rise to the additional constraint on ${\cal A}$ that is given by 
\begin{equation}
\Phi \overline \Phi ({\cal A}- 2 m) = 0 \, . 
\end{equation}
Putting everything together, \eqref{totSG} takes the form 
\begin{equation}
\begin{aligned}
\label{TOTO}
{\cal L}_\text{TOT} =  & 
\left( \int d^2 \Theta \, 2 {\cal E} \left( -3 {\cal R} + W_0 \right) + \hc \right) 
\\
& - \beta m \int  d^4\theta\, E \, \Gamma^2 \overline \Gamma^2 \left( 
1+ 
\sqrt{-\det\left(\eta_{ab}+ \frac{1}{m} \check f_{ab}\right)} 
\right) , 
\end{aligned}
\end{equation}
where we have defined $\check f_{cd} = \frac{\rmi}{4}\sigma_{cd\gamma}{}^\alpha\varepsilon^{\gamma\beta}\left( {\cal D}_\alpha \check W_\beta + {\cal D}_\beta \check W_\alpha\right)+\hc$.\footnote{Notice that, 
to truncate the gauge vector in \eqref{TOTO} we only have to make it pure gauge by setting 
$\check V = i S -i \overline S$, where $S$ is given by \eqref{PhiS}. 
The Lagrangian \eqref{TOTO} will then reduce to the so called de Sitter supergravity \cite{Dudas:2015eha,Bergshoeff:2015tra,Hasegawa:2015bza,Bandos:2015xnf}.} Since the lowest component of $\Gamma_\alpha$ is the goldstino, namely $\Gamma_\alpha | = \gamma_\alpha$, from the constraints \eqref{CSG} we have 
\begin{equation}
\chi_\alpha  = \gamma^\beta (\sigma^{ab} \epsilon)_{\alpha \beta} \, e_a^m e_b^n F_{mn} + \text{3-fermi terms} \, ,
\end{equation}
which allows us to express the gaugino $\chi_\alpha$ as a function of the goldstino $\gamma_\alpha$ and of the vector $v_m$. Once we write the Lagrangian \eqref{TOTO} in components and integrate out the auxiliary fields we find, up to second order in the fermions, that it is given by 
\begin{equation}
\begin{aligned}
e^{-1} {\cal L}_\text{TOT} = & - \frac12 R(e, \omega)  
+ 3 |W_0|^2  
+ \frac{1}{2} \epsilon^{klmn} (\overline \psi_k \overline \sigma_l D_m \psi_n - \psi_k \sigma_l D_m \overline \psi_n) 
\\
& 
-  W_0 \, \overline \psi_a \overline \sigma^{ab} \overline \psi_b 
-  \overline W_0 \, \psi_a  \sigma^{ab}  \psi_b 
- \beta m \left[ 1+ \sqrt{-\det 
\left(
\eta_{ab} + \frac{1}{m}  e_a^m e_b^n F_{mn} 
\right) } \right] 
\\
& + \beta m \left[ 1+ 
\sqrt{-\det 
\left(
\eta + m^{-1} F 
\right) } \right] 
\left( -2 W_0 \overline \gamma^2 
+ i D_a \gamma \sigma^a \overline \gamma 
+ i \gamma \sigma^a \overline \psi_a  
+ \hc \right)
\\
& + \frac{\beta}{4} \sqrt{-\det 
\left( \eta + m^{-1} F \right)} 
\left[ (\eta + m^{-1} F)^{-1} \right]^{ab} \times 
\\
& \qquad  \Big{\{} 2 \overline{W}_0 \gamma^2  e_a^m e_b^n F_{mn}  
+ \Big{[} i \gamma \s_a \overline \s^{cd} \overline \gamma D_b(e_d^m e_c^n F_{mn}) 
+ 2i \gamma \s_a \overline \s^{nm} D_b \overline \gamma \, F_{mn} 
\\
& \qquad \qquad \qquad \qquad  
+ i \gamma \s_b \overline \s^{nm} \overline \psi_a F_{mn} 
+ i \overline \psi_b \overline \s_a \s^{nm} \gamma F_{mn} - (a \leftrightarrow b) \Big{]} 
+ \hc \Big{\}} 
\\
& + \text{four-fermi terms} \, . 
\end{aligned}
\label{sugraaction}
\end{equation}

As a consequence of the fact that the system \eqref{Dtermgen} has been coupled to $\mathcal{N}=1$ supergravity, in the resulting action \eqref{sugraaction} the non-linear and spontaneously broken supersymmetry is gauged, while the other is explicitly broken. 
Otherwise, it is also possible to couple the original model to an ${\cal N}=2$ background, where both supersymmetries are going to be gauged. In that case, we do not expect physical differences between the supergravity completions of the Bagger--Galperin action \eqref{BG1} and of its alternative formulation \eqref{Dtermgen}. 
The reason why we did not follow this second possibility in the present work is because it was aimed at interpreting \eqref{Dtermgen} as an anti D$3$-brane, emphasising the distinction with the D$3$-brane interpretation of the original Bagger--Galperin model. Such a distinction is expected to become manifest when the D-brane is embedded into a supergravity background that preserves only half of the supersymmetry, as it happens here.

\section{Conclusions}

The main purpose of this work was to find evidence for a possible string theory origin of the new Fayet--Iliopoulos D-term introduced in \cite{Cribiori:2017laj}. Indeed, a relation to anti D3-branes was already pointed out in \cite{Cribiori:2017laj}, but in the present work we have strengthened this interpretation. The Bagger--Galperin action \cite{Bagger:1996wp} can be interpreted as the effective theory for a D3-brane with truncated spectrum. It has one preserved supersymmetry and one spontaneously broken and the bosonic sector matches with the Born--Infeld. Once we reformulated the Bagger--Galperin action, bringing it to a form where the broken supersymmetry is described by a superspace setup, we found that the supersymmetry breaking is sourced by the new Fayet--Iliopoulos D-term of \cite{Cribiori:2017laj}. Our findings therefore provide further evidence in favor of the anti D3-brane interpretation of the new Fayet--Iliopoulos D-term.

\section*{Acknowledgements}

We thank Stefanos Katmadas, Alex Kehagias, Antoine Van Proeyen and Timm Wrase for discussions. 
We are also very thankful to Hongliang Jiang for discussions related to section 2.1. 
The work of N.C. is supported by an FWF grant with the number P 30265. The work of F.F. is supported from the KU Leuven C1 grant ZKD1118 C16/16/005. The  work  of  M.T.  is  supported  by  the  FWO Odysseus Grant No.  G.0.E52.14N.

\appendix

\section{Non-linear realizations of supersymmetry} 
\label{appA}

In this appendix we review general properties of non-linear realizations of supersymmetry and we derive a series of formulas which we use throughout the paper. Various of the results we present below are new but, as they are rather technical, we decided to collect them together in order to avoid interruptions in the main part of the work. The appendix is therefore quite long, but it is meant to be self-contained and it can be read independently from the rest of the article. We recall that, in our conventions, the algebra satisfied by $\mathcal{N}=1$ superspace derivatives is
\begin{equation}
\{ D_\a , \overline D_{\dot \b} \} = -2 i \, \s_{\alpha \dot \beta}^m \p_m  \, , 
\end{equation}
all the other anticommutators being vanishing, while the supersymmetry transformations of a generic superfield $U$ can be defined in superspace as 
\begin{equation}
\delta \, U = \epsilon^\beta D_\beta \, U + \overline \epsilon_{\dot \beta} \overline D^{\dot \beta} \, U \, . 
\end{equation}

\subsection{Non-linear realizations in superspace} \label{Appendix1}

A minimal model with non-linearly realized supersymmetry is that of Volkov--Akulov \cite{Volkov:1973ix}, in which the goldstino $\lambda_\alpha$ transforms as
\begin{equation}
\label{gold-S3} 
\delta \lambda_\alpha = \epsilon_\alpha 
- i \left( \lambda \sigma^m \overline \epsilon - \epsilon \sigma^m \overline \lambda \right) \partial_m \lambda_\alpha \, . 
\end{equation}
Notice that, since we have set the supersymmetry breaking scale to unity, the fermion $\lambda_\alpha$ has unconventional mass dimension, namely $[\lambda]=-1/2$. This supersymmetry transformation can be embedded into superspace by defining a spinor superfield $\Lambda_\alpha$ that satisfies the constraints
\begin{equation}
\begin{aligned}
\label{SW}
D_\alpha \Lambda_\beta & = \epsilon_{\beta\alpha} 
+ i \, \sigma^m_{\alpha\dot\alpha}\overline\Lambda^{\dot\alpha} \p_m \Lambda_\beta \, , 
\\[.1cm]
\overline D_{\dot \alpha} \Lambda_\beta & = - i \,\Lambda^\alpha \sigma^m_{\alpha\dot\alpha} \p_m \Lambda_\beta \, , 
\end{aligned}
\end{equation}
and that has the goldstino $\lambda_\alpha$ as lowest component, namely $\Lambda_\alpha | = \lambda_\alpha$. Further properties of this spinor goldstino superfield $\Lambda_\alpha$ can be found in \cite{Samuel:1982uh,Ivanov:1978mx,Ivanov:1982bpa,Klein:2002vu}. Using $\Lambda_\alpha$ we can construct the composite real superfield 
\begin{equation}
\label{Aam}
\mathbb{A}_m^a = \delta_m^a 
- i \p_m\Lambda\sigma^a\overline\Lambda 
+ i \Lambda\sigma^a\p_m\overline\Lambda \,  
\end{equation}
and we will indicate its lowest component as $\mathbb{A}_m^a | = A_{m}^a$. 
The simplest supersymmetric action for the goldstino is given by the Volkov--Akulov Lagrangian. 
It can be recast into different forms, which can be shown to be equivalent up to field redefinitions. 
Following \cite{Cribiori:2016hdz} we choose therefore  
\begin{equation}
S^{VA} = - \int d^4 x \det [A_m^a]  
= - \int d^4 x \, d^4 \theta \, \Lambda^2 \overline\Lambda^2  \, , 
\end{equation}
where the last identity holds up to boundary terms. 

Beside for what concerns the goldstino, non-linear supersymmetry can be implemented on matter fields as well. The standard non-linear realization of supersymmetry is acting on matter scalars and fermions (only spin-1/2) as 
\begin{equation} \label{SNLsf}
\delta \phi = - i \left( \lambda \sigma^m \overline \epsilon - \epsilon \sigma^m \overline \lambda \right) \partial_m \phi \, , 
\quad 
\delta \chi_\alpha = - i \left( \lambda \sigma^m \overline \epsilon - \epsilon \sigma^m \overline \lambda \right) \partial_m \chi_\alpha \, .
\end{equation}
In order to preserve invariance under non-linear supersymmetry, spacetime derivatives need to be covariantized using $A^a_m$. 
For example $(A^{-1})_a^m\partial_m\phi$ and $(A^{-1})_a^m\partial_m\chi_\alpha$ transform as the standard non-linear realization 
of supersymmetry for the fields transforming as \eqref{SNLsf}, i.e.  
$\delta [(A^{-1})_a^n \partial_n \phi] 
= - i ( \lambda \sigma^m \overline \epsilon - \epsilon \sigma^m \overline \lambda ) \partial_m [(A^{-1})_a^n \partial_n \phi]$. The non-linear transformation of gauge vectors has the form \cite{Klein:2002vu} 
\begin{equation}
\label{DVD}
\delta u_m = - i \left( \lambda \sigma^n \overline \epsilon - \epsilon \sigma^n \overline \lambda \right) \partial_n u_m-\rmi\partial_m\left( \lambda \sigma^n \overline \epsilon - \epsilon \sigma^n \overline \lambda \right)u_n\,.
\end{equation}
Notice that this transformation differs from the standard non-linear realization \eqref{SNLsf} of scalars and fermions. 

In analogy to the case of the Volkov--Akulov goldstino $\lambda_\alpha$, the transformation rules for the scalar $\phi$ can be lifted to superspace. 
This is done by introducing a constrained superfield $C$ that satisfies the conditions 
\begin{equation}
D_\alpha C = i \, \sigma^m_{\alpha\dot\rho}\overline\Lambda^{\dot\rho} \p_m C \, , 
\quad 
\overline D_{\dot \alpha} C =  - i \,\Lambda^\rho \sigma^m_{\rho\dot\alpha} \p_m C \,  , 
\end{equation}
and has $\phi$ as the lowest component, namely $\phi = C|$. 
Similarly one constructs the non-linear realization for spin-1/2 matter fermions. 
The case of a gauge vector field, however, 
is slightly more involved and it is presented in the next subsection, 
since we prefer to introduce first other useful ingredients.

Since our interest is in describing non-linear realizations of supersymmetry by means of superspace, we would like to develop a general procedure in order to construct a standard non-linear realization out of any given superfield. This method, in particular, would apply also to the case of a gauge vector superfield. To this end, we introduce the superspace derivatives $\Pi_\alpha$ constructed in \cite{Cribiori:2016hdz}:
\begin{equation}
\label{Pdef}
\begin{aligned}
\Pi_\alpha &= D_\alpha - i \sigma^n_{\alpha \dot \alpha} \overline \Lambda^{\dot \alpha} \partial_n \, , 
\\[.1cm] 
\overline \Pi_{\dot \alpha} &= \overline D_{\dot \alpha} + i \Lambda^{\alpha} \sigma^n_{\alpha \dot \alpha}  \partial_n \equiv (\Pi_\alpha)^* \, .
\end{aligned}
\end{equation}
They satisfy the algebra 
\begin{equation} 
\label{PPproperties}
\{ \Pi_\alpha , \Pi_\beta \}  = 0 
\, , \quad 
\{ \Pi_\alpha , \overline \Pi_{\dot \beta} \} = 0 \, , 
\quad 
[ \Pi_\alpha \, ,  (\mathbb{A}^{-1})_a^{\ m} \partial_m ] = 0  \, 
\end{equation}
and their action on the spinor goldstino superfield $\Lambda_\alpha$ is
\begin{equation}
\Pi_\alpha \Lambda_\beta = \epsilon_{\beta\alpha}  \, , \quad \overline \Pi_{\dot \alpha} \Lambda_\beta = 0 \, . 
\end{equation}
From the derivatives $\Pi_\alpha$ we can then build the operator $\Pi^2 \overline{\Pi}^2$ that turns a linear realization into a standard non-linear one. In particular, for a generic superfield $U$ we have 
\begin{equation}
\begin{aligned}
D_\alpha \left(  \Pi^2 \overline{\Pi}^2 U \right) 
& = i \sigma^n_{\alpha \dot \alpha} \overline \Lambda^{\dot \alpha} \partial_n  \left(  \Pi^2 \overline{\Pi}^2 U \right) \, , 
\\[.1cm]
\overline D_{\dot \alpha} \left(  \Pi^2 \overline{\Pi}^2 U \right) 
& = - i \Lambda^{\alpha} \sigma^n_{\alpha \dot \alpha}  \partial_n   \left(  \Pi^2 \overline{\Pi}^2 U \right) \, . 
\end{aligned}
\end{equation}
This recipe is completely general. However, in most of the situations, we will need to construct a superfield transforming as the standard non-linear realization, but satisfying the additional requirement that it reduces to a given superfield $T$ when setting the goldstino to zero.
This can be done if we first multiply $T$ with goldstino superfields and then we act on it with the operator $\Pi^2 \overline \Pi^2$. 
In other words, when we consider the particular case in which $U$ is given by $\Lambda^2 \overline \Lambda^2 T$, the superfield
\begin{equation}
\label{Tdress}
\mathbb{T} = \frac{1}{16} \Pi^2 \overline \Pi^2 \left( \Lambda^2 \overline \Lambda^2 T \right) \,
\end{equation}
is then transforming as the standard non-linear realization
\begin{equation}
D_\alpha \mathbb{T} = i \, \sigma^m_{\alpha\dot\rho}\overline\Lambda^{\dot\rho} \p_m \mathbb{T} \, , 
\quad 
\overline D_{\dot \alpha} \mathbb{T} =  - i \,\Lambda^\rho \sigma^m_{\rho\dot\alpha} \p_m \mathbb{T} \, 
\end{equation}
and it satisfies the additional property
\begin{equation}
\label{Tprop}
  \mathbb{T} = T + {\cal O}\left(\Lambda , \overline \Lambda \right) \, .  
\end{equation}
The dressing of a superfield $T$ with the spinor goldstino superfield $\Lambda_\alpha$ is unique. To prove this, let us assume that, out of a given $T$, we construct two different superfields $\mathbb{T}_1$ and $\mathbb{T}_2$ of the type \eqref{Tprop} and transforming as the standard non-linear realization of supersymmetry. Since these superfields by construction differ only for terms $\mathcal{O}(\Lambda, \overline \Lambda)$ in \eqref{Tprop}, they satisfy
\begin{equation}
\label{LLTT}
\Lambda^2 \overline \Lambda^2 \mathbb{T}_1  = \Lambda^2 \overline \Lambda^2 \mathbb{T}_2 \, . 
\end{equation}
By acting with $\Pi^2 \overline \Pi^2$  we then find immediately 
\begin{equation}
\label{TT}
\mathbb{T}_1  =  \mathbb{T}_2 \, , 
\end{equation}
which proves the uniqueness of the dressing. 
In some situations the equation \eqref{LLTT} might appear in a component field form, namely  $\lambda^2 \overline \lambda^2 (\mathbb{T}_1|)  = \lambda^2 \overline \lambda^2 (\mathbb{T}_2|)$. Such an equation, however, can always be lifted back to the full superspace equation \eqref{LLTT}, because if the lowest components of two superfields match then, by supersymmetry, all the components have to match. Even in such a situation we would conclude therefore that \eqref{TT} holds.

Within this setup, a generic supersymmetric Lagrangian is of the form 
\begin{equation}
\label{SNLL}
{\cal L} = \det \left[A_m^a\right] \ {\mathbb L} | \, , 
\end{equation}
where $\mathbb{L}$ is a real superfield transforming as the standard non-linear realization of supersymmetry
\begin{equation}
\label{LL}
\Pi_\alpha {\mathbb L} = 0  \, , \quad \overline \Pi_{\dot \alpha} {\mathbb L} = 0  \, , \quad {\mathbb L} = \overline{\mathbb L} \, . 
\end{equation}
It can be given by the dressing of another superfield, but it can be more general as well. The invariance of $\eqref{SNLL}$ under supersymmetry follows in particular from 
\begin{equation}
\label{DDAL}
\begin{aligned}
D_\alpha \Big{(} \det \left[\mathbb{A}_m^a\right] \ {\mathbb L} \Big{)} 
&= i \p_n \Big{(}  \sigma^n_{\alpha\dot{\rho}} \overline \Lambda^{\dot{\rho}} \det \left[\mathbb{A}_m^a\right] 
\ {\mathbb L} \Big{)} \, , 
\\[.1cm] 
\overline D_{\dot \alpha} \Big{(} \det \left[\mathbb{A}_m^a\right] \ {\mathbb L} \Big{)} 
&= - i \p_n \Big{(} \Lambda^{\alpha} \sigma^n_{\alpha \dot \alpha}   \det \left[\mathbb{A}_m^a\right]
\ {\mathbb L} \Big{)} \, .
\end{aligned}
\end{equation}
We can prove an identity that relates non-linear supersymmetric actions to their equal form in which the integration over superspace has been carried out, namely 
\begin{equation}
\int d^4 x \,d^4\theta \, \det \left[\mathbb{A}_m^a\right] \, T=\frac1{16}\int d^4 x \, \det \left[A_m^a\right] \, \Pi^2\overline\Pi^2 T|\,, \label{IDENT}
\end{equation}
where the superfield $T$ is real but otherwise generic. Before we start proving \eqref{IDENT}, we need to generalize the equations in \eqref{DDAL} for a generic superfield which is not satisfying $\Pi_\alpha T =0$. This gives 
\be
\begin{aligned}
D_\alpha\Big( \det \left[\mathbb{A}_m^a\right] T\Big)&=\det \left[\mathbb{A}_m^a\right]  \,\Pi_\alpha T +\rmi \partial_n \Big( \sigma^n_{\alpha\dot\alpha} \overline \Lambda^{\dot \alpha} \det  \left[\mathbb{A}_m^a\right]\, T\Big) \, ,  \\
\overline D_{\dot\alpha}\Big( \det \left[\mathbb{A}_m^a\right] T\Big)&=\det \left[\mathbb{A}_m^a\right] \, \overline\Pi_{\dot \alpha} T -\rmi \partial_n \Big(\Lambda^\alpha \sigma^n_{\alpha\dot\alpha} \det \left[\mathbb{A}_m^a\right]\, T\Big)\,.\label{DPrel}
\end{aligned}
\ee
With the actions of the superspace derivatives in \eqref{DPrel} at our disposal, we can now prove \eqref{IDENT}. We begin by carrying out the superspace integral
\begin{equation}
\frac{1}{16} \int d^4 x  \, D^2\overline D^2 \det \left[\mathbb{A}\right]\,T| = - \frac{1}{16} \int d^4 x  \, D^2 \overline D^{\dot\alpha} \det \left[\mathbb{A}\right]\left( \frac1{\det \left[\mathbb{A}\right]}\overline D_{\dot\alpha} \det \left[\mathbb{A}\right]\, T\right) \Big{|}\,. \label{FirstStep}
\end{equation} 
In the brackets we constructed an operator from the superspace derivative that, by the equations in \eqref{DPrel}, can be readily replaced with $\overline\Pi_{\dot \alpha}$ plus a term containing a derivative on the $T$ superfield 
\begin{equation}
- \frac{1}{16} \int d^4 x  \, D^2 \overline D^{\dot\alpha} \det \left[\mathbb{A}\right]\Big[ \overline \Pi_{\dot\alpha}\,T-\rmi\frac1{\det \left[\mathbb{A}\right]}\partial_m\left(\Lambda^\alpha\sigma^m_{\alpha\dot\alpha}\det \left[\mathbb{A}\right]\, T\right)\Big] \Big{|}\,.
\end{equation}
The second term inside the brackets vanishes up to boundary terms, because the superspace derivates commute with the spacetime derivative. All that is left is therefore
\begin{equation}
- \frac{1}{16} \int d^4 x  \, D^2 \overline D^{\dot\alpha} \det \left[\mathbb{A}\right]\, \overline \Pi_{\dot\alpha} T|.
\end{equation}
Observe that we are now back in a situation similar to the one in \eqref{FirstStep}, where a superspace derivative is acting upon a superfield times the determinant function. We can thus exactly repeat the same procedure we did before. In the end, when all superspace derivatives are gone, or more precisely converted into $\Pi_\alpha$ derivatives, the action takes the simple form we set out to prove, i.e. 
\begin{equation}
\frac{1}{16} \int d^4 x  \, \det \left[A\right]\, \Pi^2 \overline \Pi^2 T|\,.
\end{equation}
Notice finally that, in the particular case in which $T=\Lambda^2\overline\Lambda^2 L$, with $L$ a real superfield, the identity \eqref{IDENT} becomes
\begin{equation}
\label{ALDL}
\int d^4 x \ d^4 \theta \, \Lambda^2 \overline \Lambda^2 \, L = \int d^4 x \,  \det \left[A_m^a\right] \ {\mathbb L} |    \, 
\end{equation}
and in the integrated part we get exactly the non-linear dressing of a superfield defined in \eqref{Tdress}.

\subsection{Constrained superfields} 

In the previous subsection we implemented the non-linear realization of supersymmetry by using the spinor goldstino $\Lambda_\alpha$. Another equivalent approach consists in imposing constraints on more familiar objects, as chiral or vector superfields. In this subsection therefore we introduce some ingredients of the approach to non-linear realizations in terms of constrained superfields. We will not review the complete literature, but we will only focus on some specific properties which are important for our discussions.

From the spinor goldstino superfield $\Lambda_\alpha$ we can construct a chiral superfield $\Phi$, by setting 
\begin{equation}
\begin{aligned}
\label{rocekde}
\Phi &=  - \frac14 \overline D^2 \left( \Lambda^2 \overline \Lambda^2 \right)
\\[0.1cm] 
& =  \Lambda^2 \left( 1 - i \p_m \Lambda \s^m \overline \Lambda 
- \overline \Lambda^2 \p_m \Lambda \s^{mn} \p_n \Lambda  \right)  \, , 
\end{aligned}
\end{equation}
which satisfies the properties of the constrained superfield introduced in \cite{Rocek:1978nb}, namely 
\begin{equation}
\label{ROCEK}
\Phi^2 = 0 \, , \quad \Phi \overline D^2 \overline \Phi = - 4 \Phi \, . 
\end{equation}
The superfield $\Phi$ contains therefore as independent component fields only one fermion $(D_\alpha \Phi |)$, which is the goldstino. 
Notice that 
\begin{equation} 
\label{phiL2}
\Phi \overline \Phi = \Lambda^2 \overline \Lambda^2 \, , 
\end{equation}
a property which we use interchangeably in this article.

Using the superfield $\Phi$ we can implement non linear-realizations also in the matter sector and construct any type of constrained superfield by eliminating specific matter component fields. The generic method to perform such a procedure has been presented in \cite{DallAgata:2016syy}. 
For the purposes of the present work, however, we are interested in constructing a specific constrained superfield which contains only a gauge vector as independent component. In the superspace description, a gauge vector is embedded into a real $\mathcal{N}=1$ superfield $\check V$, which takes the form \eqref{Vexp}, while its gauge-invariant field strength resides into the chiral superfield  $\check W_\alpha = W_\alpha(\check V)$ given in \eqref{WV}. These are both linear representations of supersymmetry. In the language of superfields, moreover, the abelian gauge transformation is encoded into 
\begin{equation}
\label{VVSS}
\check V \to \check V + i S - i \overline S \, ,  
\end{equation} 
where $S$ is a chiral superfield. Besides the vector field, the superfield $\check V$ would contain also a fermion (gaugino) and a real auxiliary field as independent components. One possible strategy to obtain a superfield describing only the vector consists in imposing additional constraints on $\check V$ in order to reduce the numbers of its independent components. These constraints will turn the linear realization of supersymmetry into a non-linear one. The gaugino is eliminated by the constraint \cite{Komargodski:2009rz} 
\begin{equation}
\label{FW}
\Phi \, \check W_\alpha = 0 \, , 
\end{equation}
while the auxiliary field D can be removed with \cite{Cribiori:2017ngp} 
\begin{equation}
\label{FFDW}
\Phi \overline \Phi \, D^\alpha \check W_\alpha = 0 \, . 
\end{equation}
Instead of the conventional Wess--Zumino gauge, which sets to zero the lower $\theta$-terms in $\check V$, we use the modified gauge choice \cite{Komargodski:2009rz} 
\begin{equation}
\label{SKG}
\Phi \check V = 0 \, , 
\end{equation}
which eliminates the component fields $\check V|$,  $D_\alpha \check V|$ and $D^2 \check V|$ 
and expresses them as functions of the remaining ones. 
In particular for $\check V$ we find 
\be
\check V = \frac12 \Lambda^\alpha \overline \Lambda^{\dot \alpha} [D_\alpha , \overline D_{\dot \alpha} ] \check V 
+ {\cal O}(`` {\Lambda^3} ") 
\ee
and for the fermionic $D_\alpha$ descendant we have
\begin{equation}
\label{DVLDDV}
\overline D_{\dot \alpha} \check V = \frac12 \Lambda^\alpha [D_\alpha , \overline D_{\dot \alpha} ] \check V 
+ {\cal O}(`` {\Lambda^2} ") \, . 
\end{equation}
With the symbol $`` {\Lambda^2} "$ we refer to terms which contain at least two bare $\Lambda$ or $\overline \Lambda$ superfields, namely terms of the form: $\Lambda^2 (\ldots)$, $\overline \Lambda^2 (\ldots)$, $\Lambda \overline \Lambda (\ldots)$; similarly for the terms  $`` {\Lambda^3} "$. Notice that, because of the gauge choice \eqref{SKG}, the chiral superfield $S$ entering \eqref{VVSS} has to be constrained to satisfy  
\begin{equation}
\label{PhiS}
\Phi S = \Phi \overline S \, . 
\end{equation} 
The constrained superfield $S$, moreover, satisfies the property 
\begin{equation} 
\label{PROPS}
\Pi^2 \overline \Pi^2 \left( \Lambda^2 \overline \Lambda^2 \, \p_a (S + \overline S) \right) 
=  
({\mathbb A}^{-1})_a^n \, \p_n \left[ \Pi^2 \overline \Pi^2 \left( \Lambda^2 \overline \Lambda^2 (S + \overline S) \right) \right] \, , 
\end{equation}
which is going to be used in a while, in order to recast into a standard form the gauge transformation of the dressed vector superfield we are going to construct. The proof of \eqref{PROPS} indeed is slightly involved, but it can be simplified if one multiplies both sides with $\Lambda^2 \overline \Lambda^2$ and shows that they match. Taking then into account that 
\begin{equation}
\Pi_\alpha \left\{ ({\mathbb A}^{-1})_a^n \, \p_n \left[ \Pi^2 \overline \Pi^2 \left( \Lambda^2 \overline \Lambda^2 (S + \overline S) \right) \right] \right\} 
= 0 \, , 
\end{equation}
which follows from \eqref{PPproperties}, and by using the uniqueness of the dressing, one can derive \eqref{PROPS}. 

We can now construct a non-linear representation of supersymmetry out of the vector component field, along the lines of what has been done for scalars and fermions in the previous section. In particular we now have all the ingredients at our disposal in order to generalize the transformation \eqref{DVD} to superspace and to construct the covariant field strength of the vector, appropriately dressed with goldstini and transforming as a standard non-linear realization of supersymmetry. Instead of working directly with $\check V$, which contains the vector field in the $\theta \overline \theta$-component, we define a real vector superfield as a descendant of $\check V$, namely
\begin{equation}
\label{V_a}
\check V_{\alpha \dot \alpha} =\check V_a\,\sigma^a_{\alpha \dot \alpha}= - \frac12 [D_\alpha , \overline D_{\dot \alpha} ] \check V \, , 
\end{equation} 
which has the lowest component field given precisely by $\check V_{\alpha \dot \alpha} | = v_{\alpha \dot \alpha}$. 
Under the gauge transformation \eqref{VVSS} the superfield $\check V_a$ transforms as 
\begin{equation}
\label{GTV}
\check V_a \to \check V_a + \p_a (S + \overline S ) \, . 
\end{equation}
We now dress the $\check V_a$ superfield with the goldstino superfield defining
\begin{equation}
{\mathbb V}_m = \frac{1}{16} \, {\mathbb A}_m^a \, \Pi^2 \overline \Pi^2 \left( \Lambda^2 \overline \Lambda^2 \, \check V_a \right)  \, . 
\end{equation} 
Under \eqref{GTV} this superfield transforms as 
\begin{equation}
{\mathbb V}_m 
\to 
{\mathbb V}_m +  \frac{1}{16} \, {\mathbb A}_m^a \, \Pi^2 \overline \Pi^2 \left( \Lambda^2 \overline \Lambda^2 \, \p_a (S + \overline S ) \right) \, , 
\end{equation}
which, with the use of \eqref{PROPS}, 
can 
be 
recast 
into 
\begin{equation}
{\mathbb V}_m 
\to 
{\mathbb V}_m + \frac{1}{16} \p_m \left[ \Pi^2 \overline \Pi^2 \left( \Lambda^2 \overline \Lambda^2 (S + \overline S) \right) \right]  \, ,  
\end{equation}
which is a standard gauge transformation. If we focus on the lowest component
\begin{equation}
{\mathbb V}_m | = u_m \, , 
\end{equation}
we see indeed that its supersymmetry transformation matches the standard non-linear transformation in \eqref{DVD}. Notice also that if we expand ${\mathbb V}_m$ in the superfield $\Lambda_\alpha$ we have 
\begin{equation}
\label{mVVL} 
{\mathbb V}_m = \check V_m + {\cal O}(`` {\Lambda^2} ") \, , 
\end{equation}  
which can be proved with the use of \eqref{DVLDDV}. Finally we can construct the covariant field-strength superfield for ${\mathbb V}_a$ by setting 
\begin{equation}
\label{FFAA}
{\mathbb F}_{ab} = \frac{1}{16} \Pi^2 \overline \Pi^2 \left\{ \Lambda^2 \overline \Lambda^2 \left( \p_a {\mathbb V}_b - \p_b {\mathbb V}_a \right) \right\} 
= 
({\mathbb A}^{-1})_a^m ({\mathbb A}^{-1})_b^n \left( \p_m {\mathbb V}_n - \p_n {\mathbb V}_m \right) \, . 
\end{equation}

\subsection{The alternative spinor goldstino superfield} \label{AGS}

In this subsection we discuss an alternative superfield description of the goldstino together with its relationship with the previous ones. As shown in \cite{Samuel:1982uh} it is possible to map, with a field redefinition, the Volkov--Akulov goldstino $\lambda_\alpha$ to another goldstino field $\gamma_\alpha$, whose supersymmetry transformation is still non-linearly realized, but chiral
\begin{equation}
\delta \gamma_\alpha = \epsilon_\alpha - 2i \gamma \sigma^m \overline\epsilon \partial_m\gamma_\alpha.
\end{equation}
To embed this new goldstino field $\gamma_\alpha$ into superspace, we define a superfield $\Gamma_\alpha$ which satisfies the conditions 
\begin{equation}
\begin{aligned}
\label{ALT}
D_\alpha \Gamma_\beta & = \epsilon_{\beta\alpha}  \, , 
\\[.1cm]
\overline D_{\dot \alpha} \Gamma_\beta & = - 2 i \,\Gamma^\rho \sigma^m_{\rho\dot\alpha} \p_m \Gamma_\beta \, , 
\end{aligned}
\end{equation} 
together with $\Gamma_\alpha| = \gamma_\alpha$. It is possible then to relate the goldstino superfields $\Lambda_\alpha$ and $\Gamma_\alpha$ directly in superspace, by means of a superfield redefinition of the form
\begin{equation}
\label{gamma}
\Gamma_\alpha = -2 \frac{D_\alpha \overline D^2 \Lambda^2 \overline \Lambda^2}{D^2 \overline D^2 \Lambda^2 \overline \Lambda^2}  \, . 
\end{equation}

In the following we are going to study the relations among Lagrangians formulated in terms of $\Lambda_\alpha$ and those given in terms of $\Gamma_\alpha$. To this purpose, let us assume that we have a nilpotent chiral superfield $X$ which satisfies
\begin{equation}
X^2 = 0 \, . 
\end{equation}
This constrained superfield is similar to the $\Phi$ introduced in the previous subsection, with the difference that it satisfies just the nilpotent constraint. The important point is that such a superfield is related to $\Gamma_\alpha$ by
\begin{equation}
\label{GammaX}
\Gamma_\alpha = -2 \frac{D_\alpha X}{D^2 X} \, ,
\end{equation}
which means that we can always perform a field redefinition between $D_\alpha X$ (and $D^2 X$) and $\Gamma_\alpha$. 
Taking advantage of the nilpotency of $X$ we have also
\begin{equation}
X = - \frac{D^\alpha X D_\alpha X}{D^2 X} 
= - \frac{D^\alpha X}{D^2 X} \frac{D_\alpha X}{D^2 X} D^2 X 
= - \frac14 \Gamma^2 D^2 X \, , 
\end{equation} 
which implies 
\begin{equation}
\label{motorhead}
\overline D_{\dot \alpha} \left(  \Gamma^2 D^2 X \right) = 0 \, . 
\end{equation}
A supersymmetry Lagrangian can be constructed in terms of $X$ as 
\begin{equation}
\label{goldLGX}
{\cal L}  = F^X + \overline F^X  \, , \quad F^X =  -\frac14 D^2 X | \, ,
\end{equation} 
which is equivalent, up to total derivatives, to another one of the form $\int d^2 \theta X$.  We prefer however to consider \eqref{goldLGX} explicitly, because we want to keep control on how it changes under supersymmetry. In particular, as a consequence of 
\begin{equation}
\label{DDXX}
\overline{D}_{\dot \alpha} D^2 X = -2i \partial_{\rho \dot \alpha} \left( \Gamma^\rho D^2 X \right) \, ,  
\end{equation}
the Lagrangian \eqref{goldLGX} will transform as 
\begin{equation}
\label{SSYY}
\delta {\cal L} = \delta ( F^X + \overline F^X  ) 
= -2 i \partial_a \left( \gamma \sigma^a \overline \epsilon \, F^X \right) 
- 2 i \partial_a \left( \overline \gamma \sigma^a \epsilon \, \overline F^X \right) \, .
\end{equation}
We would like to stress here that the superspace embedding of the supersymmetry transformation of $F^X$ in \eqref{SSYY} is uniquely fixed to be given by \eqref{DDXX} and vice versa.

We are now in a position to postulate a relation between actions invariant under non-linearly realized supersymmetry which are constructed with the superfield $\Gamma_\alpha$ and those which are constructed with the $\Lambda_\alpha$. We start by acting on the bosonic superfield $D^2 X$ and deriving 
\begin{equation}
\label{XXB}
X \overline X \, D^2 X = - 4 X \overline X \,  B \, , 
\end{equation}
which essentially defines $B$ as the part of $D^2 X$ which does not contain bare goldstini. We can also express \eqref{XXB} in a component form, namely 
\begin{equation}
\label{ggFb}
\gamma^2 \overline \gamma^2 \, F^X = \gamma^2 \overline \gamma^2 \, b \, , 
\end{equation} 
where $B| = b$. We stress that \eqref{ggFb} and \eqref{XXB} are exactly one and the same equation, once written in superspace and once in component form. In particular, $b$ is the part of $F^X$ which does not contain any bare goldstino $\gamma_\alpha$, namely 
\begin{equation}
F^X = b + {\cal O}(\gamma , \overline \gamma) \, . 
\end{equation}
Notice that, as a consequence of the relation \eqref{gamma}, we can also express \eqref{XXB} as 
\begin{equation}
\Lambda^2 \overline \Lambda^2 D^2 X = - 4 \Lambda^2 \overline \Lambda^2 B \, , 
\quad 
F^X = b + {\cal O}(\lambda , \overline \lambda) 
\end{equation} 
and therefore we have 
\begin{equation}
\begin{aligned}
\label{D2X=D2X}
D^2 X 
& = - \frac14 D^2 \left( \Gamma^2 D^2 X \right) 
= \frac{1}{16} D^2 \left( \Gamma^2 D^2 X \, \overline D^2 \overline \Gamma^2 \right) 
\\
& = \frac{1}{16} D^2 \overline D^2 \left( \Gamma^2 \overline \Gamma^2 D^2 X \right) 
= \frac{1}{16} D^2 \overline D^2 \left( \Lambda^2 \overline \Lambda^2 D^2 X \right) \\
&=- \frac{1}{4} D^2 \overline D^2 \left( \Lambda^2 \overline \Lambda^2 B \right) \, .
\end{aligned}
\end{equation}
To prove the above formula one has to take into account \eqref{motorhead}, which is used when going from the first to the second line. From \eqref{D2X=D2X} we see that if we know the form of $\Lambda^2 \overline \Lambda^2 B$, just from the properties of $X$ we can derive $D^2 X$ which is the complete Lagrangian. We can obtain then the following chain of equalities
\begin{equation}
- \frac14 D^2 X | = \int d^4 \theta \, \Lambda^2 \overline \Lambda^2 B 
= \int d^4 \theta \, \Lambda^2 \overline \Lambda^2 \det \mathbb{A}_m^a \,  \mathbb{B} 
= \det \mathbb{A}_m^a \,  \mathbb{B} \,  | 
= \det A_m^a \, {\cal B} 
\, , 
\end{equation} 
where 
\begin{equation}
\label{bbB}
\mathbb{B} = \frac{1}{16} \Pi^2 \overline \Pi^2 \left( \Lambda^2 \overline \Lambda^2 \, B \right) \, , 
\end{equation}
and 
\begin{equation}
\label{calBG}
{\cal B} = \mathbb{B} | \, . 
\end{equation}

We summarize here the results of this subsection and we recast them in a purely component field formulation. From our considerations we can conclude that, if we have a Lagrangian of the type
\begin{equation}
{\cal L} = F^X + \overline F^X \, , 
\end{equation}
which transforms under supersymmetry as 
\begin{equation}
\delta {\cal L} = \delta ( F^X + \overline F^X  ) 
= -2 i \partial_a \left( \gamma \sigma^a \overline \epsilon \, F^X \right) 
- 2 i \partial_a \left( \epsilon \sigma^a \overline \gamma \, \overline F^X \right) \, 
\end{equation}
then up to boundary terms it takes the form 
\begin{equation}
\label{LBB}
{\cal L} =  \det A_m^a \left( {\cal B} + \overline{\cal B} \right) \, , 
\end{equation} 
where ${\cal B}$ is defined as the dressing of $F^X$ under the nonlinear realization induced by the Volkov--Akulov goldstino $\lambda_\alpha$, namely 
\begin{equation}
\label{calBL}
{\cal B} \equiv e^{\delta_\epsilon} F^X |_{\epsilon=-\lambda} . 
\end{equation}
Since all possible dressings of a field under non-linear realizations of supersymmetry are equivalent, as we proved in \eqref{TT}, then  ${\cal B}$ defined in \eqref{calBL} and in \eqref{calBG} are the same object.

\subsection{Composite component fields} \label{Appendix3} 
We devote this subsection to the study of non-linear realizations of supersymmetry at the level of component fields. 
We first define $v_\epsilon^m$ such that the standard non-linear realization of the Volkov--Akulov goldstino \eqref{gold-S3} takes the form 
\begin{equation}
\delta\lambda_\alpha=\epsilon_\alpha-v_\epsilon^m\partial_m\lambda_\alpha\,,\qquad v_\epsilon^m=\rmi\lambda\sigma^m\overline\epsilon-\rmi\epsilon\sigma^m\overline\lambda\,.
\end{equation}
The specific supersymmetry variation of the goldstino allows for the construction of a composite object $\hat H$, from an arbitrary field $H$, that transforms to only the derivative term
\begin{equation}
\delta_\epsilon \hat H=-v^m_\epsilon\partial_m \hat H\,.\label{SNLtransf}
\end{equation}
The composite field $\hat H$ is the projection of the superfield ${\cal H}$, which is a generic superfield with lowest component $H$, on the hypersurface $\theta=-\lambda$. Explicitly $\hat H$ is obtained from $H$ by acting with the supersymmetry operator and projecting the parameter
\begin{equation}
\hat H={\cal H}|_{\theta=-\lambda}=(e^{\delta_\epsilon}H)|_{\epsilon=-\lambda}\,.\label{dressedH}
\end{equation}
To proof that \eqref{dressedH} transforms as \eqref{SNLtransf} we introduce the transformations
\begin{equation}
\tilde\delta_\epsilon=\delta_\epsilon+v^m_\epsilon\partial_m\,,\qquad \left[\tilde\delta_\epsilon,\tilde\delta_\eta\right]=0\,\label{newalgebra}
\end{equation}
and we notice that, since
\begin{equation}
v^m_\epsilon|_{\epsilon=-\lambda}=0\,,\qquad (\delta^k_\epsilon v_\epsilon^m)|_{\epsilon=-\lambda}=0 \label{propv}
\end{equation}
with $k\in\mathbb{N}$, we can write equivalently
\begin{equation}
\hat H=(e^{\delta_\epsilon}H)|_{\epsilon=-\lambda}=(e^{\tilde\delta_\epsilon}H)|_{\epsilon=-\lambda}\,.
\end{equation}
The variation of $\hat H$ under the newly defined operator is
\begin{align}
\tilde\delta_\eta {\hat H}&=\tilde\delta_\eta(e^{\tilde\delta_\epsilon}H)|_{\epsilon=-\lambda}\nonumber\\
&=\tilde\delta_\eta\left(\sum_{k=0}^\infty\frac{\tilde\delta_\epsilon^k H}{k!}\right) \Big{|}_{\epsilon=-\lambda}\nonumber\\
&=\left(\sum_{k=0}^\infty\frac{\tilde\delta_\eta\tilde\delta_\epsilon^k H}{k!}\right)\Big{|}_{\epsilon=-\lambda}
+\left(\sum_{k=1}^\infty\frac{\tilde\delta_{-\eta}\tilde\delta_\epsilon^{k-1} H}{(k-1)!}\right)\Big{|}_{\epsilon=-\lambda}\nonumber\\
&=0\,,
\end{align}
where we used that $\tilde \delta_{-\eta}=-\tilde \delta_\eta$.
We now list properties of the operator $e^{\delta_\epsilon}|_{\epsilon=-\lambda}$ and give the accompanied proofs.
\begin{itemize} 
\item {\bf Property 1:} ${(e^{\delta_\epsilon} \lambda)|_{\epsilon=-\lambda}=0\,.}$\\ The proof is almost immediate. We provide the few steps needed
\begin{align}
(e^{\delta_\epsilon}\lambda)|_{\epsilon=-\lambda}&=(e^{\tilde\delta_\epsilon}\lambda)|_{\epsilon=-\lambda}\nonumber\\
&=\left(\sum_{k=0}^\infty\frac{\tilde\delta_\epsilon^k\lambda}{k!}\right)\Big{|}_{\epsilon=-\lambda}\nonumber\\
&=\lambda+\epsilon|_{\epsilon=-\lambda}
+\left(\sum_{k=2}^\infty\frac{\tilde\delta_\epsilon^{k}\lambda}{k!}\right)\Big{|}_{\epsilon=-\lambda}\nonumber\\
&=0\,.\label{prop1}
\end{align}
\item {\bf Property 2:} ${(e^{\delta_\epsilon} \lambda H)|_{\epsilon=-\lambda}=0\,.}$\\
The proof is similar to the previous one. We again give the few steps needed
\begin{align}
(e^{\delta_\epsilon}\lambda H)|_{\epsilon=-\lambda}&=(e^{\tilde\delta_\epsilon}\lambda H)|_{\epsilon=-\lambda}\nonumber\\
&=\left(\sum_{k=0}^\infty\frac{1}{k!}\,\lambda\tilde\delta^k_\epsilon H\right)\Big{|}_{\epsilon=-\lambda}+\left(\sum_{k=1}^\infty \frac{1}{k!}\,k\epsilon\tilde\delta_\epsilon^{k-1}H\right)\Big{|}_{\epsilon=-\lambda}\nonumber\\
&=\left(\lambda\sum_{k=0}^\infty\frac{1}{k!}\,\tilde\delta^k_\epsilon H\right)\Big{|}_{\epsilon=-\lambda}+\left(\epsilon\sum_{k=0}^\infty \frac{1}{k!}\,\tilde\delta_\epsilon^{k}H\right)\Big{|}_{\epsilon=-\lambda}\nonumber\\
&=0\,.\label{prop2}
\end{align}
\item {\bf Property 3:} ${e^{\delta_\epsilon}\hat H|_{\epsilon=-\lambda}=\hat H.\,}$\\
This is a consequence of $\delta_\epsilon \hat H|_{\epsilon=-\lambda}=0$, which follows from \eqref{SNLtransf} and \eqref{propv}.
\item {\bf Property 4:} ${e^{\delta_\epsilon}\left(GH\right)|_{\epsilon=-\lambda}=\hat{G}\hat H\,.}$\\
Since
\begin{equation}
\delta_\epsilon \hat{G}|_{\epsilon=-\lambda}=\delta_\epsilon\hat H|_{\epsilon=-\lambda}=0\,,
\end{equation}
we can be write the product of $\hat{G}$ and $\hat H$ as
\begin{equation}
\hat{G}\hat H=e^{\delta_\epsilon}\left(\hat{G}\hat H\right)|_{\epsilon=-\lambda}\,.
\end{equation}
The second property in \eqref{prop2} leads to the final result
\begin{equation}
e^{\delta_\epsilon}\left(\hat{G}\hat H\right)|_{\epsilon=-\lambda}=e^{\delta_\epsilon}\left(GH\right)|_{\epsilon=-\lambda}\,.
\end{equation}
\end{itemize}
As pointed out, the composite fields defined in this subsection are projections of superfields on a hypersurface $\theta=-\lambda$. The alignment of the field in the non-linear supersymmetry direction makes it transform in a very specific way. Even though the transformation is now devoid of a differential interpretation, we can restore the superfield description by lifting the composite field into superspace
\begin{equation}
    \mathbb{H}=e^{\delta_\theta}\hat H\,.\label{LiftSS}
\end{equation}
 Because the dressing is uniquely determined by the non-linear supersymmetry algebra, the superfield in \eqref{LiftSS} is equal to the dressed analogue in \eqref{Tdress}, which is used throughout the paper. 
 The superspace operators \eqref{Tdress} and \eqref{LiftSS} obey the properties 1 - 4 as listed above, but lifted to superspace.

\section{Bagger--Galperin action with standard Fayet--Iliopoulos term} 
\label{appB}

In this appendix we investigate the consequences of adding a standard Fayet--Iliopoulos D-term to the Bagger--Galperin action, as in \eqref{BGFI}. Notice that, because of \eqref{DDD}, the standard Fayet--Iliopoulos term is invariant under both the linear and the non-linear supersymmetry. Using the tools we have developed in this work, we can recast the Lagrangian \eqref{BGFI} into a form where the non-linear realization of supersymmetry is manifest. In particular, since the standard Fayet--Iliopoulos term will contribute to the supersymmetry breaking, we can perform a splitting of the type \eqref{splitting} and we can then follow the procedure outlined in the previous parts of the Appendix for dressing the superfields. Following this strategy, the Lagrangian \eqref{BGFI} becomes 
\begin{equation}
\begin{aligned}
\label{KGBpre}
{\cal L}_\text{BG + FI} = & 
- \beta m \det[A_m^a] \, \sqrt{1+\frac{1}{2m^2}({\cal F}^2 
-2 {\cal D}^2)-\frac1{16m^4}({\cal F}\tilde {\cal F})^2} 
\\
&+ \beta m \det[A_m^a]  - \sqrt 2 \, \tilde \xi \, \det[A_p^e] \, {\cal D} 
\\
& + 2 \sqrt 2 i \, \tilde \xi \, \det[A_p^e] \, \epsilon^{abcd} \, [(A^{-1})_a^{\ n} \, \p_n  \lambda] \, \sigma_b \, [(A^{-1})_c^{\ k} \, \p_k \overline \lambda] \, (A^{-1})_d^{\ m} \, u_m  \, . 
\end{aligned}
\end{equation}
The equation of motion of the auxiliary field ${\cal D}$ is
\begin{equation}
\label{DDDD}
{\cal D} = \frac{\sqrt 2 m \tilde \xi}{\sqrt{\beta^2 + 2 \tilde \xi^2}} \sqrt{-\det \left( \eta + m^{-1} {\cal F} \right)} \,  
\end{equation}
and the on-shell Lagrangian takes the form
\begin{equation}
\label{KGB}
\begin{aligned}
    {\cal L}_\text{BG + FI} = &  \,  
 \beta m \det[A_m^a] - \det[A_m^a] \, m \, \sqrt{\beta^2+2\tilde \xi^2} \, 
\sqrt{-\det \left( \eta_{ab} + \frac 1m {\cal F}_{ab} \right) } 
\\
& + 2 \sqrt 2 i \, \tilde \xi \, \det[A_p^e] \, \epsilon^{abcd} \, [(A^{-1})_a^{\ n} \, \p_n  \lambda] \, \sigma_b \, [(A^{-1})_c^{\ k} \, \p_k \overline \lambda] \, (A^{-1})_d^{\ m} \, u_m  \, . 
\end{aligned}
\end{equation} 
This Lagrangian has a series of properties, which are reported below. 
\begin{itemize} 
\item Supersymmetry is still partially broken, albeit rotated because of \eqref{DDDD} \cite{Antoniadis:2008uk,Kuzenko:2009ym}. 
\item The pure bosonic sector matches with that of the Bagger--Galperin model, up to a normalization and an additive constant. 
\item The limit $\tilde \xi \to 0$ should be taken with care because the field redefinitions we performed to derive \eqref{KGBpre} are implicitly assuming $\tilde \xi \ne 0$. 
\item It presents a crucial difference with respect to the Bagger--Galperin Lagrangian, which is related to the presence of the term in the second line. As we have already explained, this term is responsible for the gauging of the R-symmetry, once we couple the theory to supergravity. 
\end{itemize}

\section{The alternative Bagger--Galperin action in tensor calculus} 

For completeness we would like to present here the alternative form of the Bagger--Galperin action in the formalism of \cite{FVP}. The superconformal version of the Lagrangian \eqref{DtermgenSG} reads
\begin{equation}
\begin{aligned}
{\cal L}_{\overline{\text{BG}}}&=-\frac{a}{4b}\left[\overline\lambda P_L \lambda\right]_F
-2a\left[\phi^0\overline{\phi^0}\frac{w^2\overline w^2}{T(\overline w^2)\overline T(w^2)}(V)_D\right]_D 
\\
&\quad +ab\left[(\phi^0\overline{\phi^0})^2\frac{w^2\overline w^2}{T(\overline w^2)\overline T(w^2)}\left(1+\frac{1}{4b^2}\mathfrak{f}_{ab}{\mathfrak f}^{ab}-\sqrt{-\det\left(\eta_{ab}+\frac1b \mathfrak{f}_{ab}\right)}\right)\right]_D \, , 
\end{aligned}
\end{equation}
with 
\begin{equation}
w^2=\frac{\overline\lambda P_L \lambda}{(\phi^0)^2}\,,
\quad \overline w^2=\frac{\overline\lambda P_R \lambda}{(\overline{\phi^0})^2}\,,
\quad \mathfrak{f}_{ab}=\frac{\phi^0}{8\overline{\phi^0}}C^{\alpha\gamma}\left(\gamma_{ab}\right)_{\gamma}{}^\beta\left({\cal Q}_\alpha\frac{\lambda_\beta}{(\phi^0)^2}+{\cal Q}_\beta\frac{\lambda_\alpha}{(\phi^0)^2}\right)\,.\label{wandf}
\end{equation}
The ${\cal Q}$ operator in \eqref{wandf} is defined from the supersymmetry transformations
\begin{equation}
\delta {\cal C} =\epsilon^\alpha {\cal Q}_\alpha {\cal C}  \, , 
\end{equation}
for a generic component field ${\cal C}$. The spinors are all written in the Majorana representation. The Weyl and chiral weights $(w,c)$ of a gauge multiplet $V=\left\{v_\mu,\lambda, D\right\}$ are $(0,0)$. It follows that the weights of the gaugino $P_L\lambda$ and auxiliary field $(V)_D\equiv D$  are respectively $(3/2,3/2)$ and $(2,0)$. 
The superconformal algebra imposes restrictions on the existence of superconformal multiplets \cite{Ferrara:2016een}. For example the weights of the fermionic bilinear $w^2$, which are $(1,1)$, allow the chiral projection $T(\overline w ^2)$, with weights $(2,2)$, to be well defined.
The superconformal field strength $\mathfrak{f}_{ab}$ has weights $(0,0)$ and is constructed such that it can serve as the lowest component of a superconformal primary multiplet.

After conformal gauge fixing $(\phi^0=\overline{\phi^0}=\kappa^{-1}= 1/ \sqrt{8 \pi G} \,, P_L\Omega^0=P_R\Omega^0=0$), the dimensionless parameters $a$ and $b$ are related to the dimensionful $\beta$ and $m$ in the following way
\begin{equation}
    a=\kappa^2\beta\,,\quad b=\kappa^2 m\,.
\end{equation}

\end{document}